\RequirePackage{fix-cm}
\documentclass[twocolumn]{svjour3}          
\smartqed  
\usepackage{graphicx}
\usepackage{booktabs}
\usepackage{amsmath}
\usepackage{amssymb}
\usepackage{float}
\usepackage{cuted}
\usepackage{enumitem}
\usepackage[numbers,sort&compress]{natbib}
\journalname{Eur. Phys. J. C}
\begin{document}

\title{Null Geodesics, Thermodynamics, Weak Gravitational Lensing, and Black Hole Shadow Characteristics of a Frolov Regular Black Hole with Constraints from EHT Observations
}

\titlerunning{Frolov Regular Black Hole: Geodesics, Lensing, and Shadow}        

\author{Shubham Kala         \and
        Hemwati Nandan \and
        Kush Maithani \and
        Saswati Roy \and 
        Amare Abebe 
}


\institute{Shubham Kala \at
              The Institute of Mathematical Sciences\\
              C.I.T. Campus, Taramani, Chennai-600113\\
              Tamil Nadu, India\\
              \email{shubhamkala871@gmail.com}           
           \and
           Hemwati Nandan \at
              Department of Physics, Hemvati Nandan Bahuguna Garhwal Central University, Srinagar Garhwal, Uttarakhand 246174, India\\
              Centre for Space Research, North-West University, Potchefstroom 2520, South Africa\\
              \email{hnandan@associated.iucaa.in}
            \and
            Kush Maithani \at
            Department of Physics, Hemvati Nandan Bahuguna Garhwal Central University, Srinagar Garhwal, Uttarakhand 246174, India\\
            \email{kushm.mai98@gmail.com}
            \and
            Saswati Roy \at
            Department of Physics, National Institute of Technology, Agartala, Tripura, 799046, India\\
            \email{sr.phy2011@yahoo.com}
            \and
            Amare Abebe \at
            Centre for Space Research, North-West University, Potchefstroom 2520, South Africa\\
            National Institute for Theoretical and Computational Sciences (NITheCS), South Africa\\
            \email{Amare.Abebe@nithecs.ac.za}
            }
           
\date{Received: date / Accepted: date}

\maketitle

\begin{abstract}
In this paper, we investigate the properties of null geodesics, thermodynamics, gravitational lensing, and black hole shadows in the vicinity of a static regular Frolov black hole. By analyzing the trajectories of null geodesics, we investigate the bending of light in weak field regimes. The black hole shadow is studied in detail, with constraints on its parameters derived from observational data of the EHT collaboration. Further, we examine shadow images under a spherically symmetric accretion flow and compute the energy emission rate to understand the black hole radiation characteristics. The obtained results demonstrate how the Frolov black hole differs from well-known black hole solutions, such as Schwarzschild, Reissner-Nordström, and Hayward black holes. This study provides new insights into the impact of modified non-rotating regular black hole metrics on observational signatures. The findings have significant implications for future astrophysical observations and the testing of alternative gravity theories.
\end{abstract}
\tableofcontents
\section{Introduction} \label{section1}
Black holes (BHs) remain one of the most fascinating and enigmatic predictions of general relativity (GR), continuously challenging our understanding of spacetime and fundamental physics \cite{Einstein:1916vd,Misner:1973prb,chandrasekhar1998mathematical,Hartle:2003yu}. Among the diverse categories of BHs, the regular BHs are important for several reasons, both theoretical and observational. These BHs are models in which the singularity, typically associated with infinite density and curvature at the center of a classical BH, is replaced by a more regular structure. In 1968, Bardeen introduced the first regular BH solution, which avoids the singularity typically found at the center of a classical BH \cite{bardeen1968non}. After Bardeen, Hayward extended the concept of regular BHs in 1998 by presenting a new non-singular solution that incorporated a specific form of matter to avoid the singularity at the center \cite{Hayward:2005gi}. Following, Florov in 2016 further generalized regular BH solutions and presented a non-singular model for a charged BH, which obeys the limiting curvature condition \cite{frolov2016notes}. Furthermore, various non-singular BH solutions have been reported in both GR and alternative theories of gravity, which can be found in the references \cite{Nicolini:2008aj,Lemos:2008cv,Guendelman:2009vz,Castro:2011fm,Bambi:2016wdn,Olmo:2016tra,Chinaglia:2017uqd,Lamy:2018zvj,Brahma:2020eos,Kumar:2020yem,Han:2022rsx,Bambi:2023try}.\\ 
The study of BH thermodynamics gives an interrelationship among the laws of thermodynamics and BH mechanics. Bekenstein and Hawking first proposed the relation between BHs and their thermal properties \cite{bekenstein2020black,Hawking:1975vcx}. Thermodynamic characteristics of non-singular BHs have also garnered significant attention. Jingyun et al. \cite{Man:2013hpa} investigated the thermodynamic properties of Bard-een BHs, including local temperature, heat capacity, and off-shell free energy. This study also explores the effects of varying the BH charge on its existence and stability. Maluf et al. studied the thermodynamics of a class of regular BHs within the framework of the generalized uncertainty principle \cite{Maluf:2018lyu}. The thermodynamics of Bardeen BH in the context of regularized 4D Einstein–Gauss–Bonnet gravity were studied in greater detail by Kumar et al. \cite{Kumar:2020uyz}. The effects of thermal fluctuations on the thermodynamics of the modified Hayward BH were studied, with the reference provided here \cite{Pourhassan:2016qoz}. A more in-depth discussion of the thermodynamics of the Hayward BH can be found here \cite{Molina:2021hgx}. More recently, the thermodynamics of Florov BH in the presence of quintessence has been reported by Gohain et.el. \cite{Gohain:2024piy}. The thermodynamics of various non-singular BHs have been studied within the framework of alternative theories of gravity. Some recent studies can be referred to for a detailed analysis \cite{Aros:2019quj,Singh:2022xgi,Cadoni:2022chn,Odintsov:2023qfj,Misyura:2024fho}. In this work, our aim is to correct the first law of thermodynamics for regular BHs using the methodology proposed by Meng et al. \cite{Ma:2014qma}, which has been extensively adopted in the field. For more details, refer to the references cited \cite{Sajadi:2017glu,Singh:2020xju,Rodrigues:2022qdp,Sudhanshu:2024wqb,Kumar:2024qon}.\\
Gravitational lensing (GL), one of the most intriguing phenomena predicted by GR, arises when light is deflected as it passes near a compact object, such as a BH. Classical methods rooted in GR laid the groundwork for later advances in the calculation of deflection angles. Refsdal’s seminal work introduced the concept of multiple images in gravitational lensing, demonstrating how gravitational fields can bend light to produce distinct images of a background source \cite{Refsdal:1964yk}. Virbhadra and Ellis resurrected the theoretical study of BH lensing, leading to the widely accepted Virbhadra-Ellis (VE) lens equation \cite{Virbhadra:1999nm}. Virbhadra further advanced the field by analyzing relativistic images in GL, developing techniques to constrain the compactness of massive objects, and more recently, extended this work and coined the term photohole, referring to a compact object enclosed within a photon surface or photon ring \cite{Virbhadra:2008ws,Virbhadra:2022ybp}. In 2008, Gibbons and Werner revolutionized the study of gravitational deflection by introducing a geometrical and topological approach based on the Gauss-Bonnet (GB) theorem, applied to static and spherically symmetric spacetimes \cite{Gibbons:2008rj}. Werner later extended this framework to rotating and stationary spacetimes using Randers-Finsler geometry \cite{Werner:2012rc}. Most calculations employing the GB theorem traditionally assume the weak-field limit, with the source and receiver positioned at infinite distances from the lensing object. However, practical astrophysical scenarios typically involve finite distances. The Gibbons-Werner method enabled researchers to account for finite distances between the source, lens, and receiver in analyzing light deflection. Ishihara et al. applied the GB theorem to investigate finite-distance deflection in static and spherically symmetric spacetimes, considering both weak- and strong-field limits \cite{Ishihara:2016vdc,PhysRevD.95.044017}. Building on this, Ono et al. introduced the generalized optical metric method, extending its application to stationary, axisymmetric, and asymptotically flat spacetimes \cite{PhysRevD.98.044047}. Arakida also made significant contributions to the study of finite-distance light deflection in static and spherically symmetric spacetimes \cite{arakida2018light}. Recently, Huang et al. derived a simplified formula for the deflection angle by refining the geometric expression of the GB theorem double integral \cite{Huang:2023bto}. More recently, an analytic generalization of the weak field deflection angle (WDA) is derived by utilizing the current non-asymptotically flat generalization of the GB theorem \cite{Pantig:2024kqy}. Over the past decade, numerous studies have adopted to analyze Gl, enriching our understanding of compact objects and the astrophysical implications of gravitational lensing. For a comprehensive review of the field, refer to the relevant literature \cite{Iyer:2009wa,Ghaffarnejad:2014zva,Jusufi:2017drg,Abdujabbarov:2017pfw,Ovgun:2018oxk,Ovgun:2018prw,Uniyal:2018ngj,Pang:2018jpm,Kumar:2019pjp,Wang:2019cuf,Islam:2020xmy,Kala:2020prt,Kala:2020viz,Okyay:2021nnh,Gao:2021luq,Belhaj:2021lpi,Kala:2021ppi,Atamurotov:2022wsr,Belhaj:2022vte,Pantig:2022gih,Parbin:2023zik,Ditta:2023ccf,Feng:2024zxi,Pantig:2024ixc,Kala:2024fvg,Vishvakarma:2024icz,Kala:2022uog,Kala:2025iri}.\\
The BH shadow is a dark region appearing against the luminous background of infalling material, typically from an accretion disk. It forms due to the extreme deflection of light near the event horizon, as predicted by GR. Some of these photons escape and travel to distant observers, creating the observed shadow boundary \cite{synge1966escape}. The concept of BH shadow was first introduced by James Bardeen in 1973, when he studied how a BH silhouette would appear to a faraway observer \cite{bardeen1973proceedings}. Since then, the fundamental equations governing BH shadows have been well established. Advancements in theoretical physics, numerical simulations, and observational techniques have significantly enhanced our understanding of BHs shadow. This progress reached a milestone in 2019 with the groundbreaking observation of the supermassive BH $M87$ by the Event Horizon Telescope (EHT), providing direct empirical evidence of BH shadows. \cite{Akiyama_2019,akiyama2022first}. Given the growing interest in this field, numerous studies have explored factors influencing the BH shadow shape and size, investigating potential deviations from GR \cite{Kumar:2018ple,Wang:2018prk,Haroon:2018ryd,Ayzenberg:2018jip,Ovgun:2018tua,Cunha:2018cof,Shaikh:2018lcc,Ma:2019ybz,Bisnovatyi-Kogan:2019wdd,Medeiros:2019cde,Konoplya:2019goy,Ovgun:2019jdo,Zhu:2019ura,Stuchlik:2019uvf,Zhang:2020xub,Dey:2020bgo,Belhaj:2020okh,Ghosh:2020ece,Jusufi:2020cpn,Zeng:2020dco,Kumar:2020hgm,Omwoyo:2021uah,Papnoi:2021rvw,Bronzwaer:2021lzo,Guo:2021bwr,Zhang:2021pvx,Yang:2022btw,Meng:2022kjs,Belhaj:2022rmc,Ye:2023qks,Hamil:2023neq,Mafuz:2023qxn,Guzman-Herrera:2023zsv,Heydari-Fard:2023ent,Vishvakarma:2023csw,Olmo:2023lil,Tan:2023ngk,Erices:2024lci,Liu:2024soc,Sanchez:2024sdm,Fernandes:2024ztk,Chernov:2024xwn}. These efforts continue to provide deeper insights into the nature of strong gravity and the fundamental physics governing BHs.\\
Since the discovery of the Frolov regular BH in 2016, its shadow and gravitational lensing properties have been investigated for the rotating case by Kumar et al. in 2019 \cite{Kumar:2019pjp}. More recently, these properties were studied in the presence of quintessence matter by Gohain et al. \cite{Gohain:2024piy}. However, to date, no studies on optical properties have been reported for the static case of this particular BH. Therefore, the present work focuses on the static case, providing a new avenue for exploring the optical properties of astrophysical BHs.\\
The paper is organized as follows: Section \ref{section2} explores the horizon structure, geodesic equations, effective potential, and the behavior of null geodesics. The study of the deflection angle in the weak field limit is reported in Section \ref{section4}. In Section \ref{section3}, we conduct a detailed analysis of the BH's thermodynamic properties, including Hawking temperature, heat capacity, and entropy. Section \ref{section5} focuses on the BH shadow and the constraints on its parameters based on observations of Sgr A*, along with the shadow images in a spherically symmetric accretion flow. In Section \ref{section6}, we compute the energy emission rate and, finally, we present our conclusions in Section \ref{section7}.
\section{Geodesics of Frolov Black Hole Spacetime} \label{section2}

\subsection{Spacetime \& Horizon Structure}
\begin{figure}[htbp]
	\begin{center}
        {\includegraphics[height=7cm, width=0.48\textwidth]{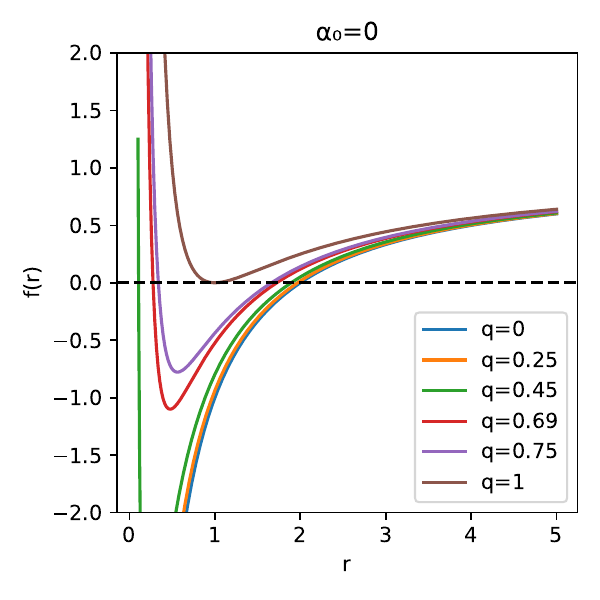}} 
        {\includegraphics[height=7cm, width=0.48\textwidth]{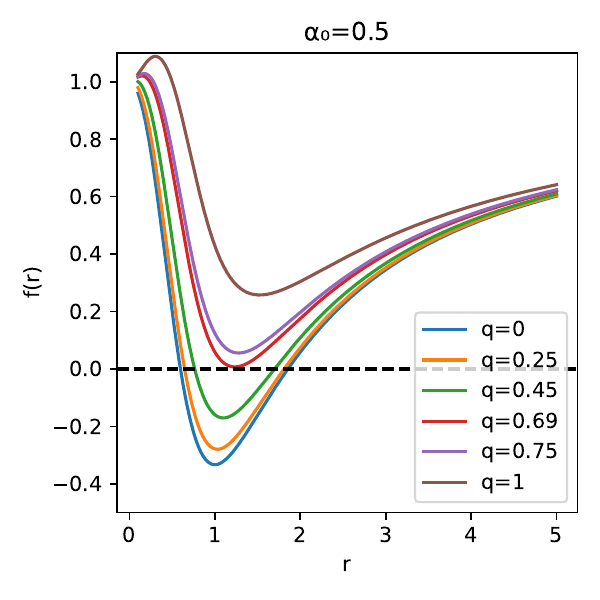}}
        {\includegraphics[height=7cm, width=0.48\textwidth]{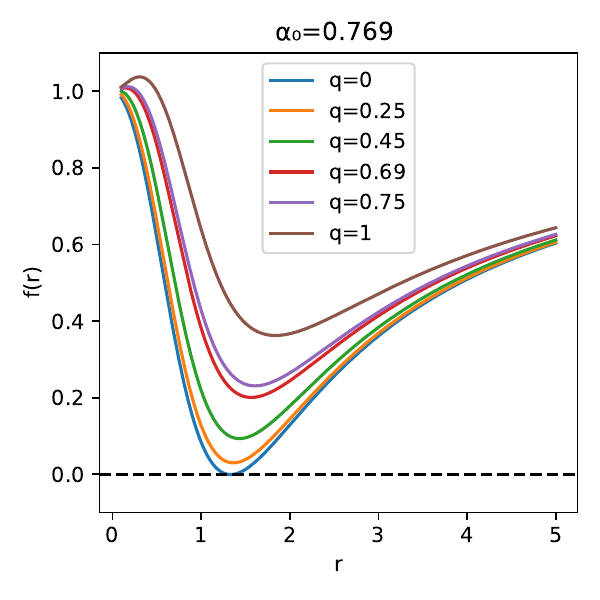}}
	\end{center}
	\caption{The variation of metric function $f(r)$ with radial distance for distinct values of $q$ and $\alpha_{0}$.} \label{Horizon}
\end{figure}

The spacetime geometry of a Frolov regular BH can be given as \cite{frolov2016notes}
\begin{equation}\label{eq_metric}
    ds^{2} = -f(r) dt^{2} + \frac{1}{f(r)} dr^{2} + r^{2} d\theta^{2} + r^2\sin^{2}\theta d\phi^{2},
\end{equation}
where
\begin{equation}
    f(r) = 1-\frac{(2Mr-q^{2})r^{2}}{r^{4}+(2Mr+q^{2})\alpha_{0}^{2}}.
\end{equation}
Here, $M$ denotes the mass of BH and $q$ is the charge parameter. The parameter $\alpha_{0}$ represents the Hubble length which is related to an effective cosmological constant i.e., $\Lambda = 3/\alpha_{0}^{2}$. The Hubble length characterizes a universal hair and is bounded by the following inequality,
\begin{equation}\label{eq_alpha}
    \alpha_{0} \leq M\sqrt{16/27}.
\end{equation}
Satisfying this constraint leads to significant quantum gravity effects. The Frolov BH reduces to the Hayward BH at $q=0$, and if we set $\alpha_{0}=0$ it simplifies to the Reissner-Nordstr\"{o}m (RN) BH. At $q=0$ and $\alpha_{0}=0$ the Frolov BH reduces to a well-known Schwarzschild BH solution in GR. The variation of the metric function $f(r)$ with radial distance for different values of $q$ and $\alpha_{0}$ is depicted in Fig. \ref{Horizon}. We begin our horizon analysis by considering the case $\alpha_0 = 0$, which corresponds to the RN BH. As the charge parameter \( q \) increases from the Schwarzschild limit ($q=0$), the RN BH initially exhibits two distinct horizons: an inner (Cauchy) horizon and an outer (event) horizon. However, as $q$ approaches the extremal value $q=M$, these two horizons merge into a single degenerate horizon, characteristic of an extremal BH. For charge values exceeding $q=M$, the horizons disappear, leaving a naked singularity, which challenges the cosmic censorship conjecture. When $\alpha_0 = 0.5$, we observe that the system has two horizons for \( q < 0.69 \), a single horizon at $q = 0.69$, and no horizon for \( q > 0.69 \). For the Hayward BH, at $\alpha_0 = 0.5$, two horizons are present, similar to the behavior seen in the RN BH. However, as \( \alpha_0 \) reaches its maximum value of $\alpha_0 = 0.769$, the system transforms into a BH with only one horizon. When both $q$ and $\alpha_0$ are non-zero, the Frolov geometry significantly affects the horizon structure. The influence of \( \alpha_0 \) introduces significant changes compared to the RN BH case, transforming the horizon structure. For specific values of $q$ and $\alpha_0$, the system can exhibit multiple horizons, a single horizon, or even no horizon at all. Finally, for $\alpha_0 = 0.769$, we find that a single horizon exists at $q=0$, but no horizon is present for any \( q > 0 \), indicating a transition to a spacetime without horizons in this regime.

\begin{figure}[htbp]
	\begin{center}
        {\includegraphics[height=5cm, width=0.4\textwidth]{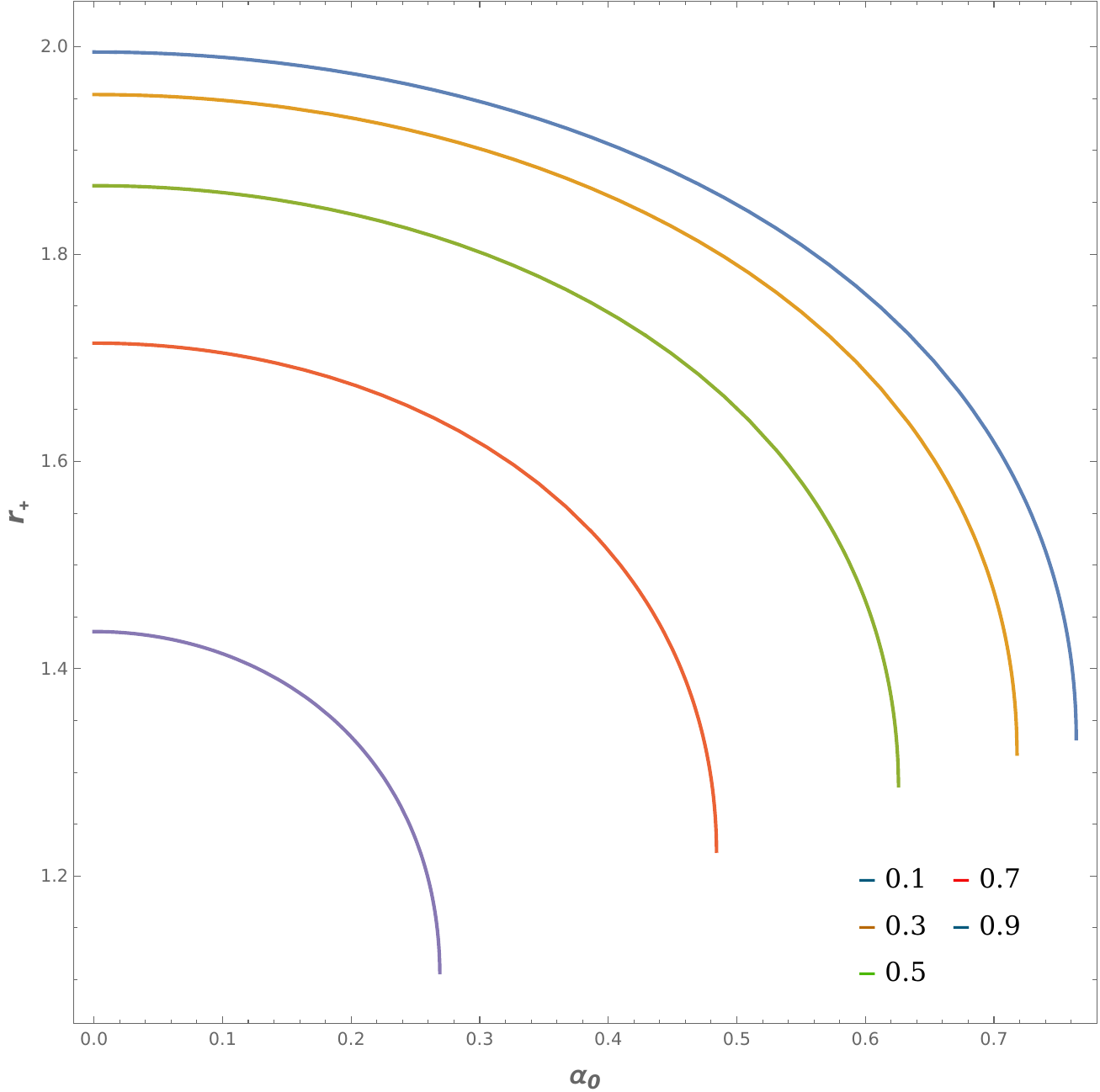}} 
        {\includegraphics[height=5cm, width=0.4\textwidth]{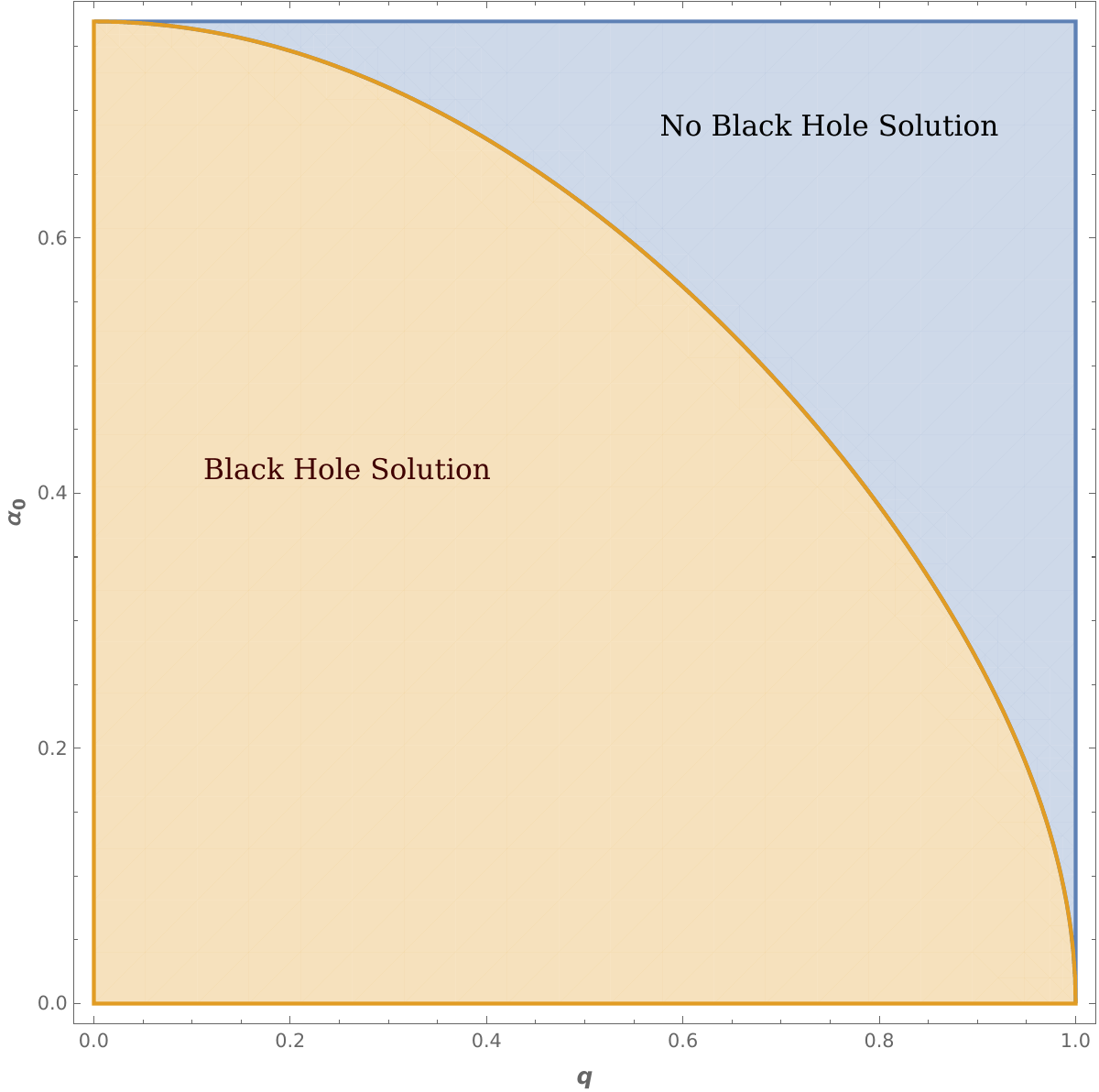}}
	\end{center}
	\caption{The variation of $r_{h+}$ with $\alpha_0$, showing that higher charge values lead to a reduction in the horizon radius (upper panel). Permissible ranges of $q$ and $\alpha_0$,  illustrating that certain combinations do not yield a valid BH solution (lower panel). } \label{fig_horizons}
\end{figure}

\noindent The horizon of an BH can be found by setting the metric function equal to zero. In our case, this results in a quadratic equation, which typically has four real roots. However, based on the variation of the metric function, we confirm that only two horizons are physically possible: the BH event horizon and the Cauchy horizon. The other two roots are imaginary and have no physical significance, so we focus on these two valid horizons. We analytically calculate the horizon radii, where $r{_h}_{+}$ represents the BH outer horizon and $r_{h-}$  represents the inner horizon which reads as,
\begin{equation}\label{eq_horizon}
    \begin{aligned}
        r_{h\pm} &= \frac{M}{2} + \frac{1}{2}\sqrt{M^{2}-\frac{2q^{2}}{3}+\Delta} 
         \pm \frac{1}{2} \left[ 2M^{2} -\frac{4q^{2}}{3}-\Delta \right. \\
        &\quad \left. - \frac{(-2M^{3}+2Mq^{2}+4M\alpha_{0}^{2})}
        {\sqrt{M^{2}-\frac{2q^{2}}{3}+\Delta}} \right]^{1/2} \;,
    \end{aligned}
\end{equation}
where
\begin{equation}\label{eq_delta}
    \Delta = \frac{2}{3} \sqrt{\mathcal{A}} \cos \Biggl\{ \frac{1}{3} \arccos \left( \frac{\mathcal{B}}{2\mathcal{A}\sqrt{\mathcal{A}}} \right) \Biggr\},
\end{equation}
with
\begin{equation}\label{eq_A}
    \mathcal{A} = 12 M^{2} \alpha_{0}^{2}  + 12 q^{2} \alpha_{ 0}^{2} + q^{4},
\end{equation}
and 
\begin{equation}
    \mathcal{B} = 144 M^{2} q^{2} \alpha_{0}^{2} + 108 M^{2}{\alpha_0^4} 
     - 72 q^{4} \alpha_{0}^{2} +2q^{6}.
\end{equation}
\begin{table*}[h]
    \centering
    \renewcommand{\arraystretch}{1.2} 
    \begin{tabular}{|c | c c c c | c c c c |}
        \toprule
        & \multicolumn{4}{c|}{\textbf{Event Horizon} ($r_h$)} & \multicolumn{4}{c|}{\textbf{Photon Sphere} ($r_p$)} \\
        \cmidrule(lr){2-5} \cmidrule(lr){6-9}
        $\alpha_0$ & $q=0.0$ & $q=0.2$ & $q=0.5$ & $q=1.0$ & $q=0.0$ & $q=0.2$ & $q=0.5$ & $q=1.0$ \\
        \midrule
        0.000 & 2.0000 & 1.9798 & 1.8660 & 1.0000 & 3.0000 & 2.9731 & 2.8229 & 1.0000 \\
        0.200 & 1.9796 & 1.9585 & 1.8388 & - - -  & 2.9820 & 2.9544 & 2.7997 & - - - \\
        0.500 & 1.8546 & 1.8266 & 1.6514 & - - -  & 2.8780 & 2.8459 & 2.6594 &- - - \\
        0.769 & 1.6562 & - - -  & - - -  & - - -  & 2.6428 & - - -  & - - -  &- - - \\
        \bottomrule
    \end{tabular}
    \caption{The values of BH event horizon ($r_{h+}$) and photon sphere ($r_{ph}$) for different values of charge ($q$) and Hubble length parameter ($\alpha_0$). The dashed lines show that for some specific parameters combination BH solution does not exist.}
    \label{tab:horizon_photon}
\end{table*}
Fig. \ref{fig_horizons} illustrates the possibility of BH solution and how the BH horizon radius varies with the charge parameter $q$ and the Hubble length parameter $(\alpha_{0})$ . As  $q$ increases, the horizon radius decreases, indicating that a higher charge reduces the BH's size. Notably, only specific values of $q$ and $\alpha_0$ are permissible, and other combinations give imaginary values of the horizon radius, suggesting that the horizon is not possible at those values. For a more comprehensive numerical validation, we provide the calculated event horizon and photon sphere radii for different combinations of the parameters $\alpha_{0}$ and $q$ in Table \ref{tab:horizon_photon}.
\subsection{Null Geodesic \& Effective Potential}
To study null geodesics, we use the Euler-Lagrange me-thod as given by,
\begin{equation}
    \frac{d}{ds} \left( \frac{\partial \mathcal{L}}{\partial \dot{x}_{\mu}} \right) = \frac{\partial \mathcal{L}}{\partial x_{\mu}},
\end{equation}
where $s$ is the affine parameter of the light trajectory, a dot represents the differentiation w.r.t. $x$ and $x_{\mu}$ represents the four-velocity of the light ray. The Lagrangian corresponding to the Frolov BH metric is given by,
\begin{equation} \label{lag}
    \mathcal{L} = \frac{1}{2} \left[ -f(r) \dot{t}^{2} + \frac{\dot{r}^{2}}{f(r)} + r^{2} (\dot{\theta}^{2} + \sin^{2}\theta \dot{\phi}^{2}) \right].
\end{equation}
From the metric it is clear that the metric coefficients do not explicitly depend on time $t$ and azimuthal angle $\phi$, there are two constants of motion which can be labeled as $E$ and $L$,
\begin{equation} \label{energy}
    E = -f(r) \dot{t}, 
\end{equation}
and
\begin{equation} \label{angm}
    L = r^{2} \sin^{2}\theta \dot{\phi}, 
\end{equation}
where $E$ and $L$ are the energy and angular momentum of the particle respectively. Further, we constrain the motion of particles on the equatorial plane and consider 
$\theta=\pi/2$ which gives $\dot{\theta}=0$. For null geodesics, consider that the Lagrangian is zero and with constants of motion obtained in Eq. \ref{energy} and Eq. \ref{angm} , the geodesic equation becomes
\begin{equation}
    \left( \frac{dr}{ds}\right)^{2}  = E^{2} - \frac{L^{2}}{r^{2}} f(r).
\end{equation}
We can obtain the radial equation as a function of $s$ and $t$ as follows,
\begin{equation} \label{eqrdot}
   \left( \frac{dr}{ds}\right)^{2} + 2 V_{eff}(r)=  E^{2},
\end{equation}
\begin{equation}
   \left( \frac{dr}{dt}\right)^{2} = \left( 1-\frac{2}{E^{2}} V_{eff}(r)  \right) f(r)^{2}.
\end{equation}
\begin{figure}[htbp] 
	\begin{center}
        {\includegraphics[height=7cm, width=0.48\textwidth]{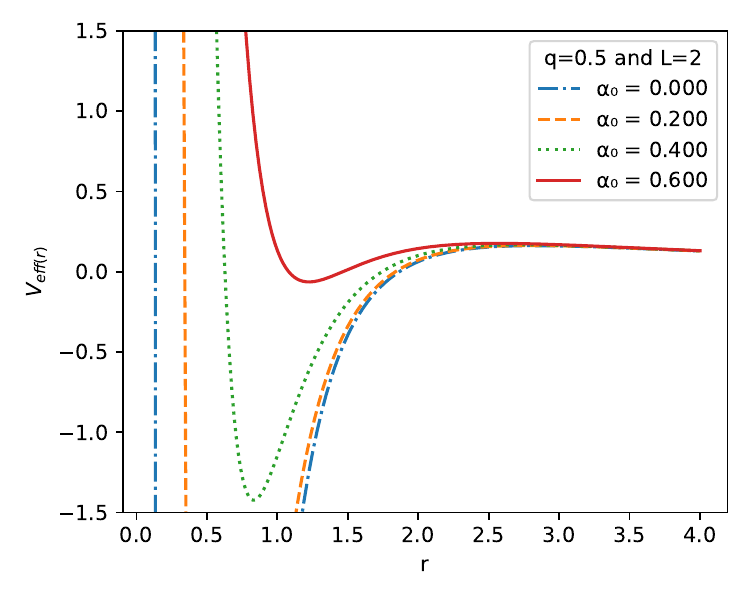}} 
        {\includegraphics[height=7cm, width=0.48\textwidth]{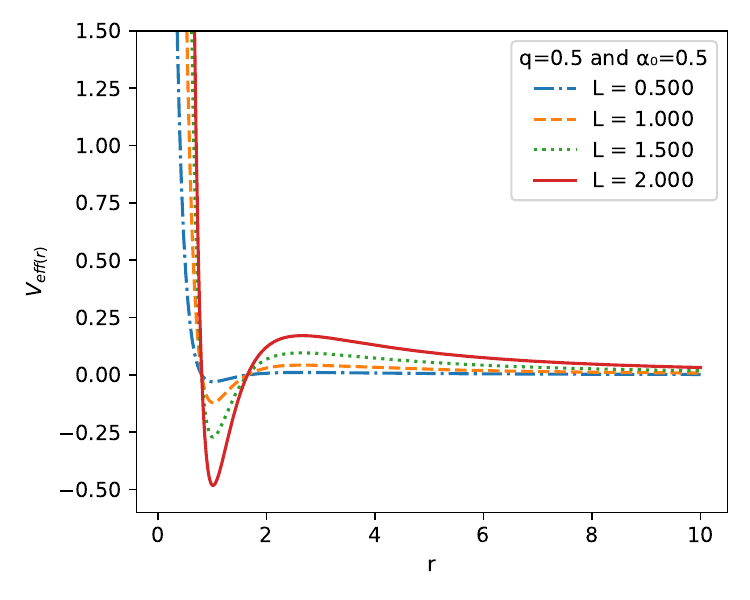}}
	\end{center}
	\caption{The variation of effective potential with radial distance for different values of $\alpha_{0}$ and $L$.} \label{efffig}
\end{figure}
Here, $V_{eff}(r)$ represents the effective potential and can be defined as,
\begin{equation} \label{effp}
  V_{eff}(r) = \left[ 1-\frac{(2Mr-q^{2})r^{2}}{r^{4}+(2Mr+q^{2})\alpha_{0}^{2}} \right] \frac{L^{2}}{2r^{2}}.
\end{equation}
\begin{figure}[H] 
	\begin{center} 
        {\includegraphics[width=0.48\textwidth]{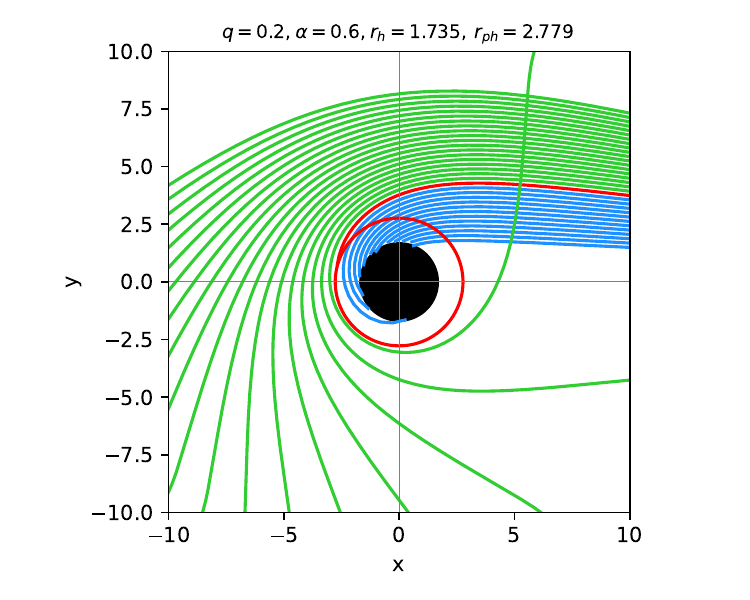}}
        {\includegraphics[width=0.48\textwidth]{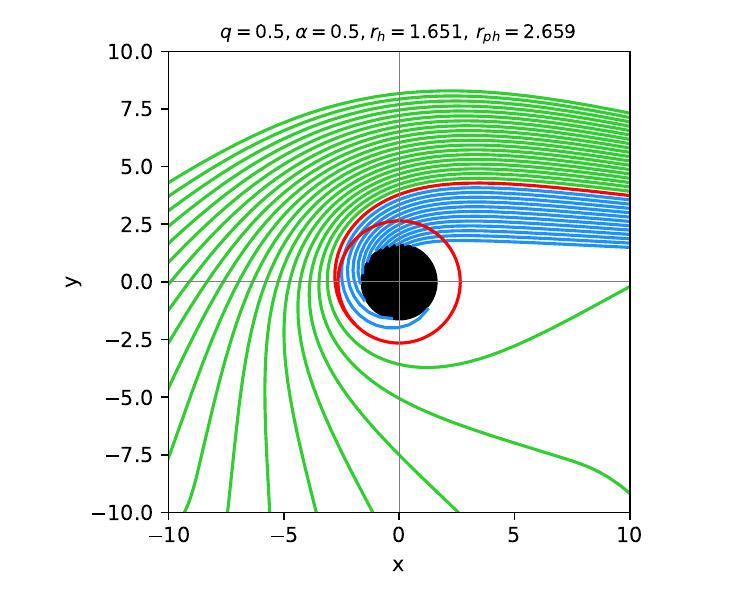}} 
        {\includegraphics[width=0.48\textwidth]{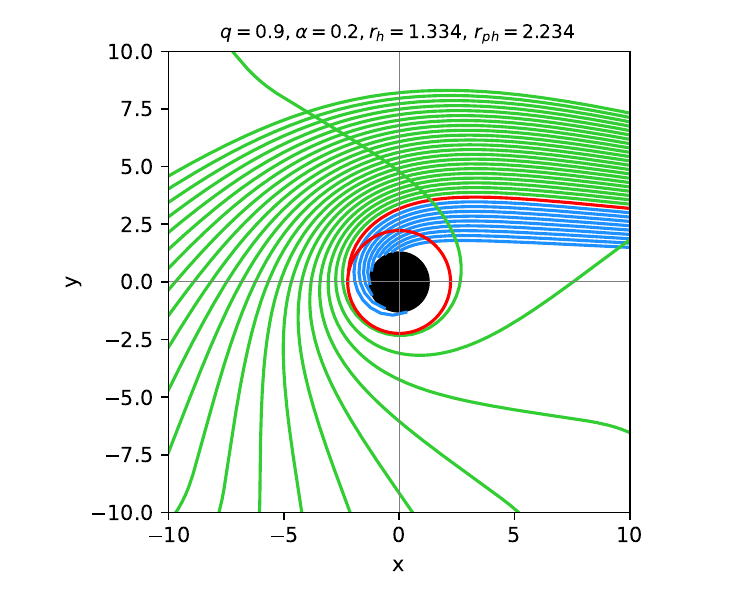}}      
	\end{center}
	\caption{Null geodesics for various parameter sets of the Frolov regular BH. The 
         blue, red, and green trajectories correspond to the absorbing, circular, and escape orbits, respectively.} \label{nullgeofig}
\end{figure}
\begin{figure}
    \centering
        {\includegraphics[width=0.48\textwidth]{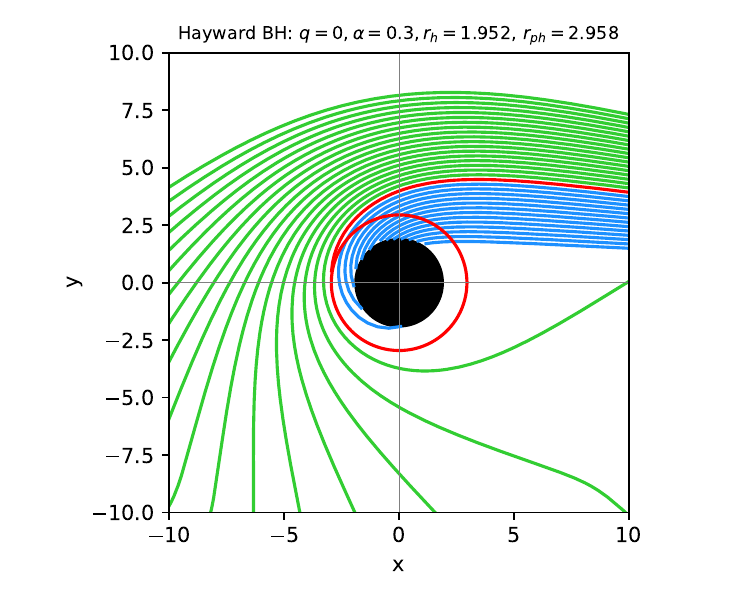}}
        {\includegraphics[width=0.48\textwidth]{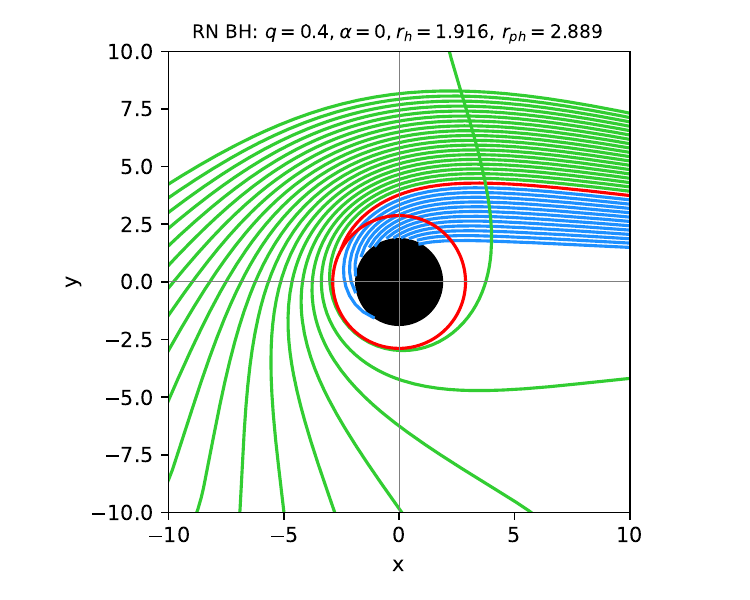}} 
        {\includegraphics[width=0.48\textwidth]{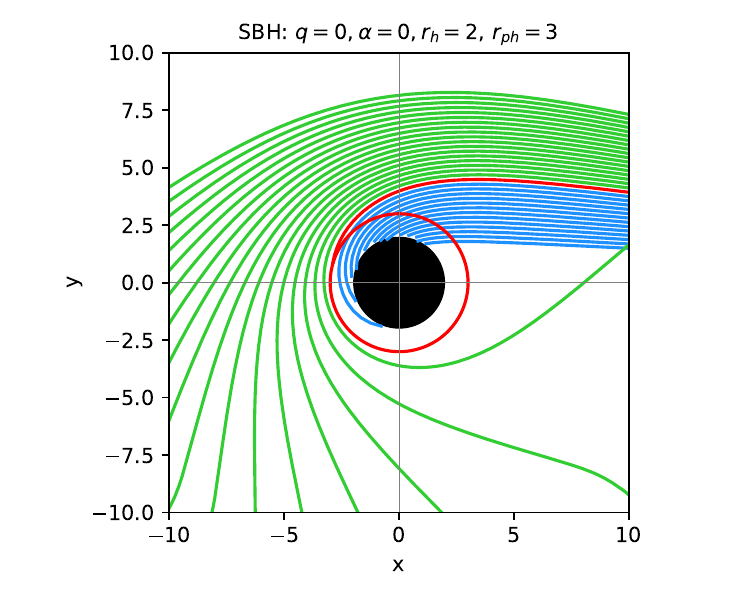}}
    \caption{For comparison, the null geodesics of well-known BHs to which the Frolov BH reduces under specific limiting cases are also illustrated.}
    \label{fig:enter-label}
\end{figure}
We can consider the ratio of angular momentum and energy as an impact parameter i.e., $b=L/E$, and the geodesic equation becomes
\begin{equation} \label{radialeqm}
   \left( \frac{dr}{ds}\right)^{2} =  L^{2} \left[ \frac{1}{b^{2}} - \frac{f(r)}{r^{2}} \right]. 
\end{equation}
From Eq. \ref{angm} and Eq. \ref{radialeqm}, we can obtain the orbit equation as,
\begin{equation}
    \left( \frac{dr}{d\phi}\right)^{2} = \left[ \frac{r^{4}}{b^{2}} - r^{2} -\frac{(2Mr-q^{2})r^{4}}{r^{4}+(2Mr+q^{2})\alpha_{0}^{2}} \right].
\end{equation}
We can make the change of variable $u=1/r$ and therefore, the orbit equation can be written as follows:
\begin{equation} \label{orbiteq}
     \left( \frac{du}{d\phi}\right)^{2} = \left[ \frac{1}{b^{2}} -u^{2} - \frac{2Mu^{3}-q^{2}u^{4}}{1+(2Mu^{3}+q^{2}u^{4})\alpha_{0}^{2}} \right].
\end{equation}
The null geodesics are determined by numerically solving the orbit equation, with the results illustrated in Fig. \ref{nullgeofig}. The blue lines correspond to trajectories with an impact parameter less than the critical value, representing absorbing trajectories. The green trajectories, with impact parameters greater than the critical value, represent escaping orbits. Circular photon orbits, corresponding to the peaks of the effective potential shown in Fig. \ref{efffig}, are highlighted as red trajectories in Fig. \ref{nullgeofig}.

\subsection{Stability of Null Geodesics via Dynamical Systems Approach}
\begin{figure}[H]
	\centering 
	    \includegraphics[height=7cm, width=0.45\textwidth, angle=0]{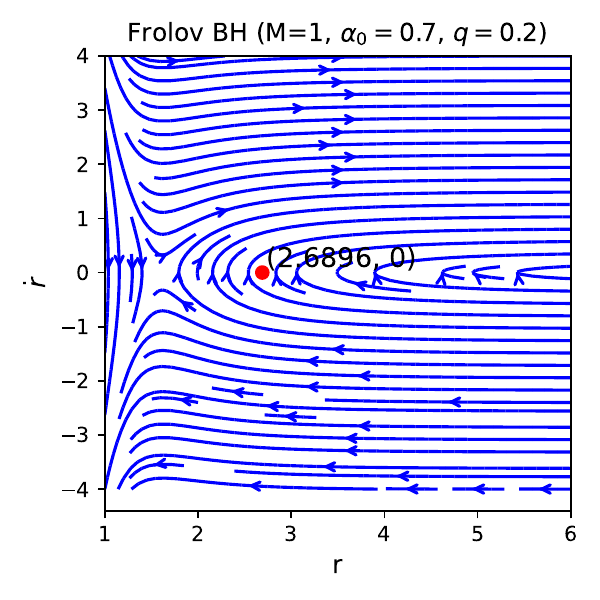}
        \includegraphics[height=7cm, width=0.45\textwidth, angle=0]{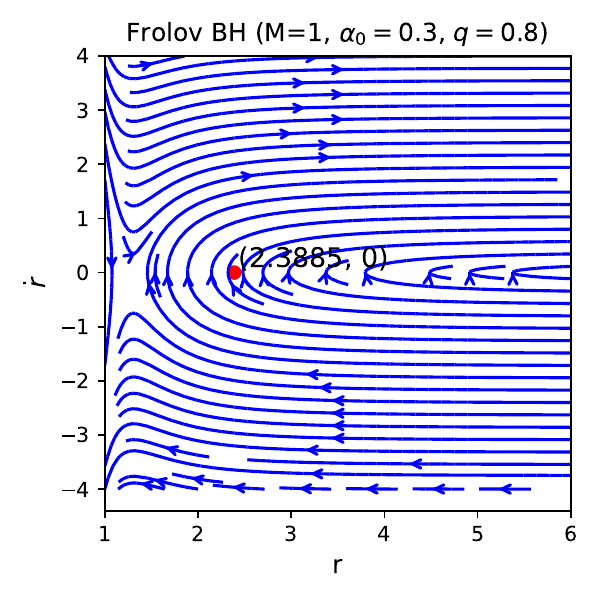}
        \includegraphics[height=7cm, width=0.45\textwidth, angle=0]{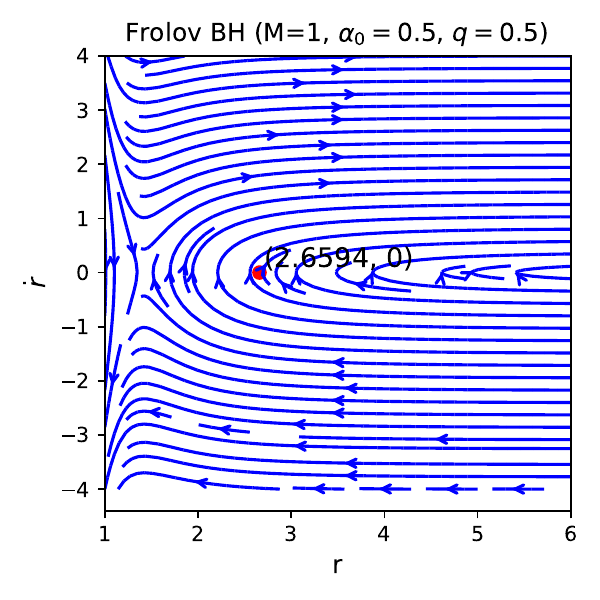}
	\caption{The graph shows the phase portrait $r$ vs $\dot{r}$ of the null geodesics for different combination of the parameter $\alpha_{0}$ and $q$. The red dot represent the location of saddle critical point.} 
	\label{stability}
\end{figure}

\begin{figure}
    \centering
       \includegraphics[height=7cm, width=0.45\textwidth, angle=0]{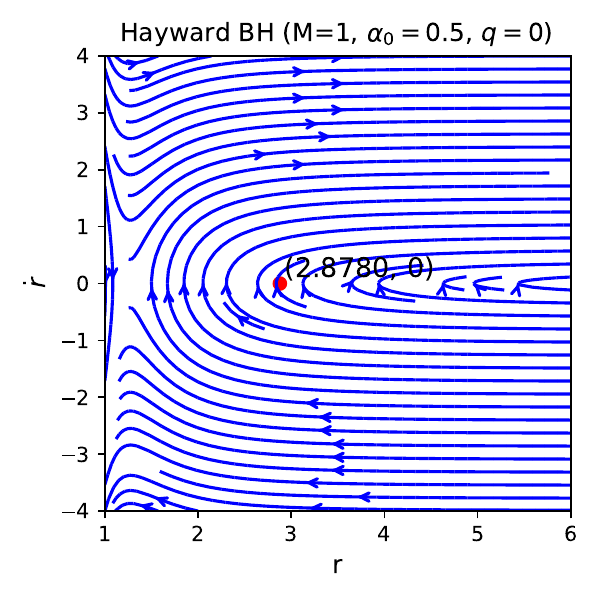}
        \includegraphics[height=7cm, width=0.45\textwidth, angle=0]{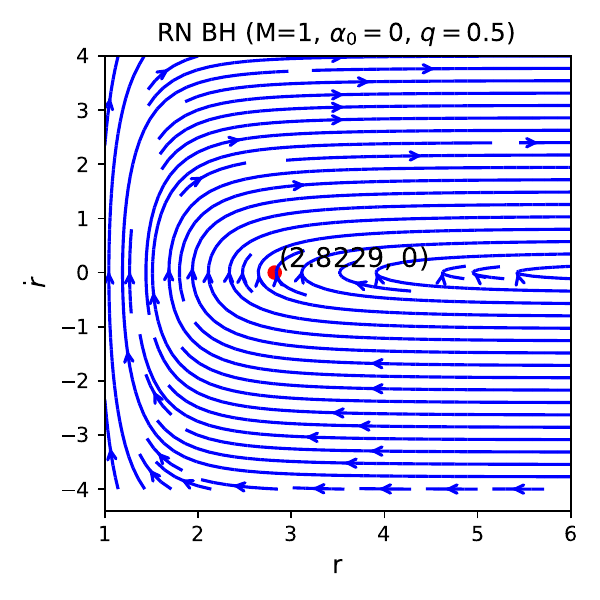}
        \includegraphics[height=7cm, width=0.45\textwidth, angle=0]{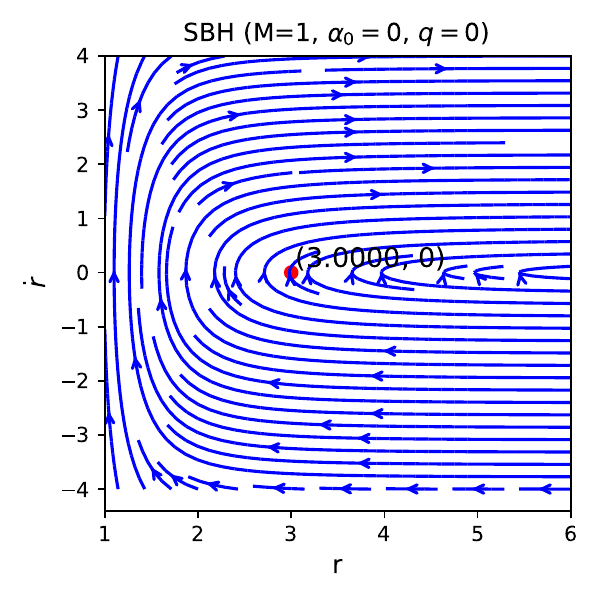}
    \caption{The graph shows the phase portrait $r$ vs $\dot{r}$ of the null geodesics for well-known BHs to which the Frolov BH reduces under specific limits.}
    \label{stability2}
\end{figure}

In this subsection, we explore the stability of null circular geodesics by constructing a dynamical framework and examining its phase space evolution in the \((r, \dot{r})\) plane \cite{goldhirsch1987stability}. The stability of these orbits is assessed by analyzing how small radial perturbations influence their trajectories. Since null geodesics satisfy the condition \(\dot{r} = 0\) at equilibrium, the radial velocity vanishes, effectively constraining the system to a reduced two-dimensional phase space defined by \(r\) and its conjugate momentum. By studying the phase flow structure in this plane, we can pinpoint the critical saddle point corresponding to the photon sphere, denoted as \((r_c, 0)\). To advance this analysis, we differentiate Eq. \ref{eqrdot} with time, eliminating $\dot{r}$ to derive the following expression:
\begin{equation}
    \ddot{r} = -\frac{dV_{\text{eff}}(r)}{dr}.
\end{equation}
Now we consider the coordinates $x_1 = \dot{r}$ and $x_2 = \dot{x_1}$, leading to the differential equations corresponding to these coordinates, which are given by,
\begin{equation} \label{JacobianM}
    \begin{aligned}
x_1 &= \dot{r}, \\
x_2 &= -\frac{dV_{\text{eff}}(r)}{dr}.
\end{aligned}
\end{equation}
The Jacobian matrix $\mathcal{J}$ corresponding to the sets of differential  Eq. \ref{JacobianM}  is given by \cite{yang2023lyapunov},
\begin{equation}
    \mathcal{J} = \begin{pmatrix}
0 & 1 \\
-V_{\text{eff}}''(r) & 0
\end{pmatrix}, 
\end{equation}
where $V_{\text{eff}}''(r)$ denotes the second derivative of the effective potential with respect to $r$. The secular equation $|\mathcal{J} - \lambda I| = 0$ yields the eigenvalue squared as
\begin{equation}
    \lambda_{L}^2 = -V_{\text{eff}}''(r). 
\end{equation}
Here, $\lambda_{L}$, known as the Lyapunov exponent, measures the average rate at which nearby trajectories in the phase space converge or diverge over time. $\lambda_{L}^2 > 0$ indicates a divergence between nearby trajectories, that is, the behavior of the system exhibits unstable directions (saddle critical points). In contrast, the condition $\lambda_{L}^2 < 0$ indicates a convergence between nearby trajectories, that is, the critical point represents a stable center. \\
The phase portraits in Fig. \ref{stability} depict the phase flow of circular null geodesics for different combinations of $\alpha_{0}$ and $q$. For all these cases, the circular geodesics corresponding to the photon sphere radius are expected to be dynamically unstable, as they reside at the peak of the effective potential. This instability implies that even an infinitesimal perturbation can cause photons to deviate from their orbit, ultimately leading them to either escape to infinity or plunge into the BH’s event horizon. If photons are located at one of these saddle points, even the smallest perturbation can cause them to deviate from their unstable circular orbits. Depending on the nature of the perturbation, the photon may either spiral into the BH or escape to infinity. This instability highlights the exponential divergence of nearby trajectories in the phase space, demonstrating the extreme sensitivity of photon orbits to infinitesimal perturbations. We observed that for lower values of $\alpha_{0}$ and higher values of $q$, the critical saddle point shifts to a lower value. This indicates a steeper effective potential, suggesting increasingly unstable conditions for photon orbits. These results can be compared with well-known BH solutions such as Schwarzschild, RN, and Hayward BH, as illustrated in Fig. \ref{stability2}. We found that the critical saddle points are located at \((3,0)\) for Schwarzschild, \((2.8780,0)\) for Hayward, \((2.8229,0)\) for RN, and \((2.6594,0)\) for Frolov BH. A lower saddle point coordinate indicates a stronger effective potential gradient, leading to greater instability. Thus, the Frolov BH exhibits the most unstable photon orbits, while the Schwarzschild BH has the least. This suggests that charge $q$ and the Hubble length parameter $\alpha_0$ enhance the sensitivity of photon orbits to perturbations, increasing their instability.
\section{Thermodynamics} \label{section3}
\subsection{Temperature}
In this section, we study the thermodynamic mechanism of the Frolov BH as given by Eq. \ref{eq_metric}.
First, we derive the thermodynamic temperature by identifying the future-directed null vector at the event horizon. This is done by considering the near-horizon geometry and applying the first law of thermodynamics. Further, we find that the temperature expression deviates from the standard result and requires a correction. This correction arises due to the modified structure of the BH, influenced by the nonlinear effect due to the Hubble length parameter. Consequently, the thermodynamic temperature of the regular BH must account for these adjustments, ensuring that the result reflects the altered near-horizon behavior.
\subsubsection{Null Vector}
\begin{figure}[H]
    \begin{center}
        {\includegraphics[height=7cm, width=0.48\textwidth]{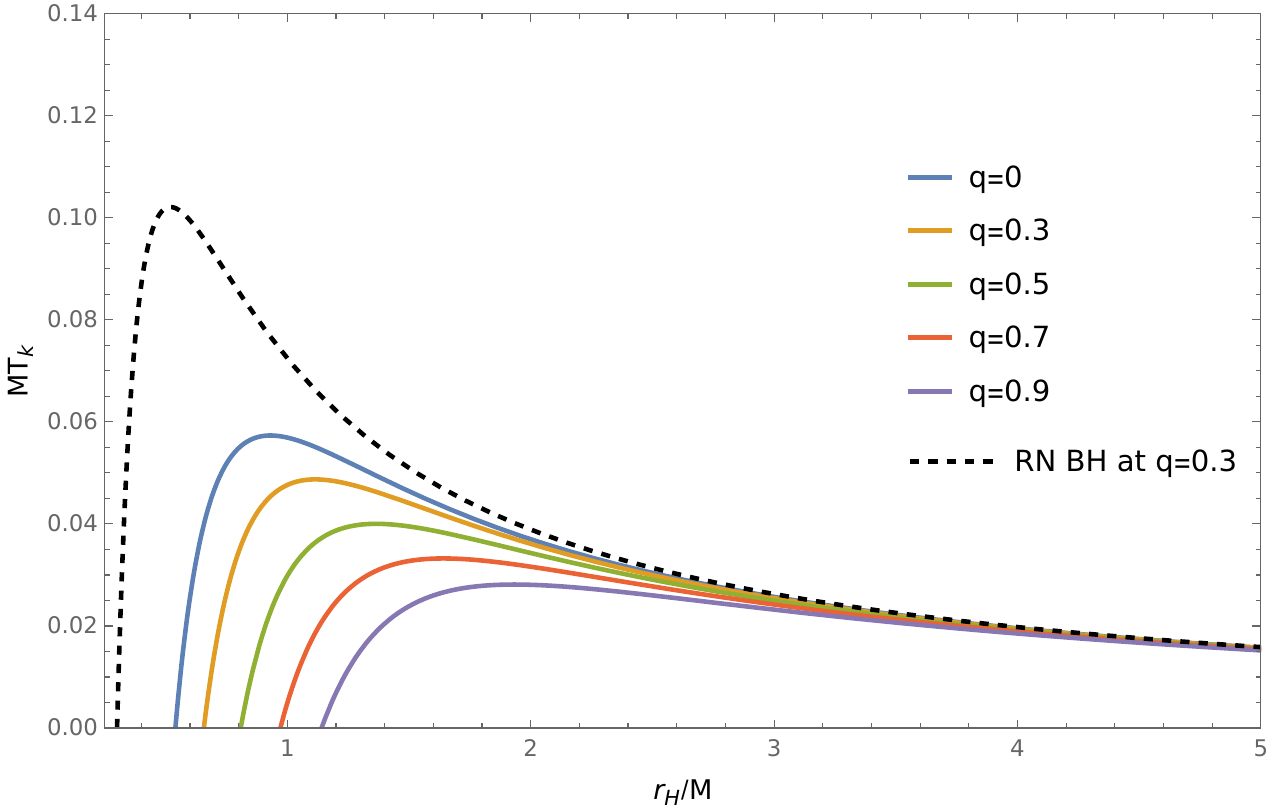} }
        {\includegraphics[height=7cm, width=0.48\textwidth]{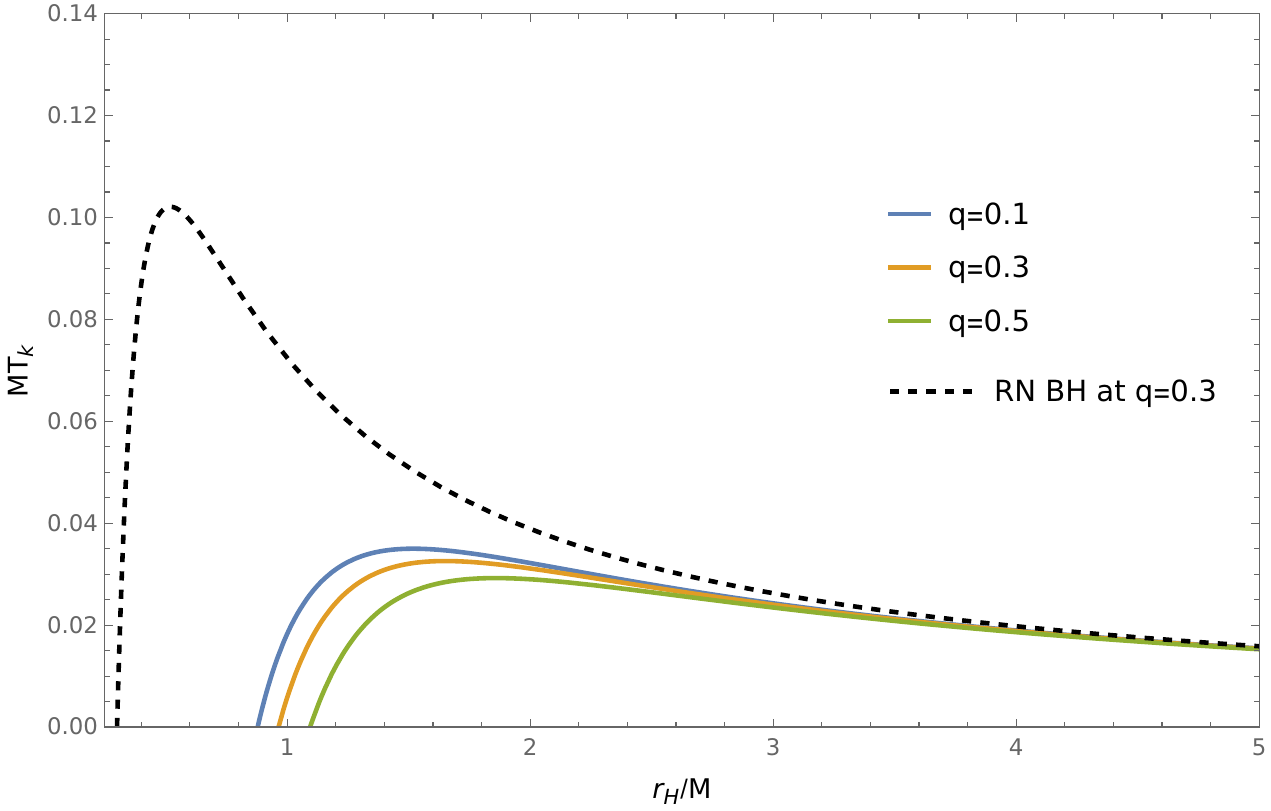} }
        {\includegraphics[height=7cm, width=0.48\textwidth]{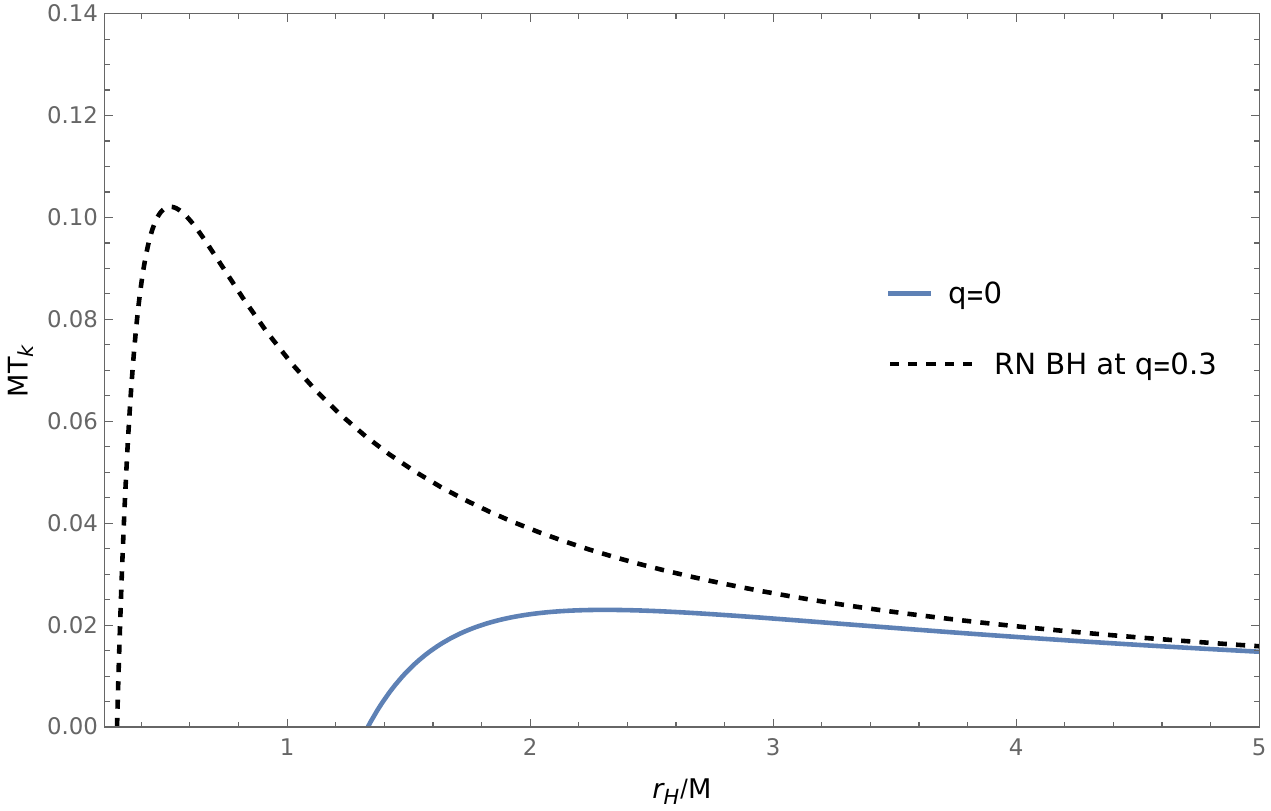} }     
        \caption{
            Variation of temperature with horizon radius for fixed $\alpha_0/M=0.3$ (upper panel), $\alpha_0/M=0.5$ (middle panel), $\alpha_0/M=0.769$ (lower panel). Here we have considered $M=1$. The values of $q$ are chosen from the permissible range (Fig. \ref{fig_horizons}). 
        } \label{thermal_plots}
    \end{center}
\end{figure}

We first define a null vector field $l$ normal at horizon, such that $l_\mu l^\mu = 0$ i.e., $g_{\mu \nu}l^\mu l^\nu = 0$. Thereafter, we choose $l$ to be a future directed null vector which is given as,
\begin{equation}
    l^{\mu} = \frac{\partial}{\partial t} + \frac{1}{2}\left( 1 + \frac{(q^2 - 2mr)r^2}{r^4 + (q^2 + 2mr)\alpha_0^2} \right) \frac{\partial}{\partial r}.
\end{equation}
The vector field $l$ must satisfies the null geodesic equations as follows \cite{Bardeen1973}:
\begin{equation}
       l^\mu \nabla_\mu l^\nu = \kappa l^\nu.
\end{equation}
\begin{equation}\label{eq4}
    \nabla_{l}l = \kappa l
\end{equation}
where $\nabla_\mu$ denotes the covariant derivative and $\kappa$ is a constant called surface gravity. $\kappa$ is constant over the horizon whose Lie derivative along $l$ vanishes, i.e., $\mathcal{L}_l \kappa = 0$.
Left hand side of Eq. (\ref{eq4}) leads to (for more details see \cite{bholes}), 
\begin{multline}
    l^\mu \nabla_\mu l^\nu =  
    \left( \frac{\alpha_0^4 q^2 - 4\alpha_0^2 q^2 r^2 - 3\alpha_0^2 r^4 - q^2r^4 + r^6}
    {2(2\alpha_0^2 q^2 + r^4)r^3} \right) \\   
    \times \left(  \frac{\partial}{\partial t} + \frac{1}{2} \left( 1 + \frac{(q^2 - 2mr)r^2}
    {r^4 + (q^2 + 2mr)\alpha_0^2} \right) \frac{\partial}{\partial r} \right).
\end{multline}
thus on comparing with the right hand side surface gravity can be expressed as,
\begin{equation}
    \kappa = \frac{\alpha_0^4 q^2 - 4\alpha_0^2 q^2 r^2 - 3\alpha_0^2 r^4 - q^2r^4 + r^6}{2(2\alpha_0^2 q^2 + r^4)r^3}.
\end{equation}
Now, the Hawking temperature can easily be calculated as
\begin{align}
        T_H &= \frac{\kappa}{2\pi} |_{r_+ },\\
            &= \frac{\alpha_0^4 q^2 - 4\alpha_0^2 q^2 r_+^2 - 3\alpha_0^2 r_+^4 - q^2r_+^4 + r_+^6}{4\pi(2\alpha_0^2 q^2 + r_+^4)r_+^3},
\end{align}
where $r_+$ denotes the horizon radius. Also, calculating temperature from surface gravity gives the same result, which is the usual way of calculating the Hawking temperature:
\begin{equation}
    T_k = \frac{\kappa}{2\pi}  = \frac{1}{2\pi} \left(\frac{1}{2}\frac{\partial f(r)}{\partial r}\right)_{r_+},
\end{equation}
\begin{equation}\label{eq9}
    T_k = \frac{\alpha_0^4 q^2 - 4\alpha_0^2 q^2 r^2 - 3\alpha_0^2 r^4 - q^2r^4 + r^6}{4\pi(2\alpha_0^2 q^2 + r^4)r^3}.
\end{equation}
This matches $T_H$ which was obtained from the null vector. On rearranging, we get
\begin{multline}
     T_k = T_{H(RN)}\left[\frac{1}{\frac{\alpha_0^2}{r^2} 
     \left( \frac{q^2}{r^2} + \frac{2m}{r} \right) + 1} \right]^2 \\ 
     + \frac{\alpha_0^2 r}{2\pi} 
     \left[  \frac{q^4 - 2mq^2r - 4m^2r^2}
     {\left(\alpha_0^2 (q^2 + 2mr)+ r^4\right)^2} \right] \;.
\end{multline}
where the term $T_{H(RN)} = \frac{mr - q^2}{2\pi r^3}$ is usual RN BH temperature. For $\alpha_0 = 0$, $T_k$ reduces to RN temperature. A comparison is shown in Fig. 
\ref{thermal_plots} against RN BH (at $q=0.3$). It is apparent that Frolov BH is a bit ``cooler'' than RN BH and the temperature is even lower for larger values of charge parameter.
Another important observation is that the Hawking temperature decreases as the charge increases, consistent with the behavior of the RN BH and for larger horizon radii $r_h$ the temperature approaches that of the RN BH.
\subsubsection{First Law}
The Hawking temperature of the BH can also be obtained by the first law of thermodynamics.
From the usual thermodynamics law $dM = TdS + PdV$,
\begin{align*}
    T_f &= \left( \frac{\partial M}{\partial S} \right)_{r_+} = 
        \frac{\partial M}{\partial r_+} \left( \frac{\partial S}{\partial r_+} \right)^{-1},\\
        &= \frac{\alpha_0^4 q^2 - 4\alpha_0^2 q^2 r^2 - 3\alpha_0^2 r^4 - q^2r^4 + r^6}{4\pi (\alpha_0 + r)^2 (\alpha_0 - r)^2}.
\end{align*}
It can be seen that the temperature $T_f$ is different from the $T_H$ obtained from the surface gravity in the previous section. The plots of $T_k$ and $T_f$ are shown in Fig. \ref{thermal_plots}.
\subsubsection{Correction Function}
The discrepancy between the Hawking temperature \( T_H \), derived from the surface gravity of the BH, and \( T_f \), obtained from the first law of thermodynamics, presents a significant issue. First law of thermodynamics $dM = TdS + PdV$ is violated if the zeroth component of  energy momentum tensor depends on the BH mass, in that case we need to modify the mass term using correction function $\mathcal{W}(r_+, \Lambda)$ given by \cite{Ma:2014qma},
\begin{equation*}
    \mathcal{W}(r_+, \Lambda) = 1 + 4\pi \int_{r_+}^{\infty} r^2\frac{\partial T_0^0}{\partial M}dr,
\end{equation*}
or in a simpler way,
\begin{equation*}
    \mathcal{W}(r_+, \Lambda) = \frac{T_k}{T_f} = \frac{(\alpha_0 + r)^2(\alpha_0 - r)^2}{(2\alpha_0 q^2 + r^4)}.
\end{equation*}
The modified mass is given by, $\mathcal{M} = \mathcal{W} (r_+, \Lambda)M$, then the modified temperature can be calculated as
\begin{align}\label{eq6}
    T'_f &= \frac{\partial \mathcal{M}}{\partial S} = \mathcal{W}(r_+, \Lambda) \frac{\partial M}{\partial S} \notag \\
         &\implies T'_f = \mathcal{W}(r_+, \Lambda) T_f = T_k.
\end{align}
is the corrected form of temperature which is one obtained from the surface gravity.
Again from (\ref{eq6}),
\begin{align}\label{eq_entropy}
    \begin{split}
        dS &= \frac{\mathcal{W}(r_+, \Lambda)}{T'_f}dM\\
         S &= \int \frac{\mathcal{W}(r_+, \Lambda)}{T'_f}\frac{\partial M}{\partial r_+}dr_+ = \pi r_+^2.
    \end{split}
\end{align}
\subsection{Heat Capacity}
\begin{figure}[H]
    \begin{center}
        {\includegraphics[width=0.48\textwidth]{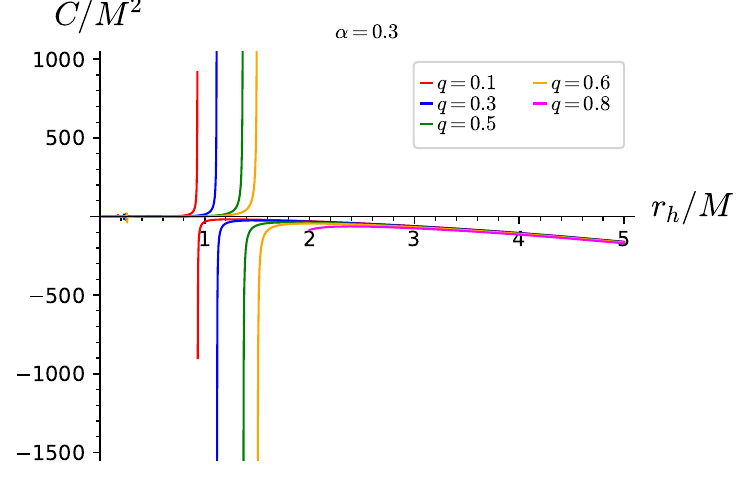} }
        {\includegraphics[width=0.48\textwidth]{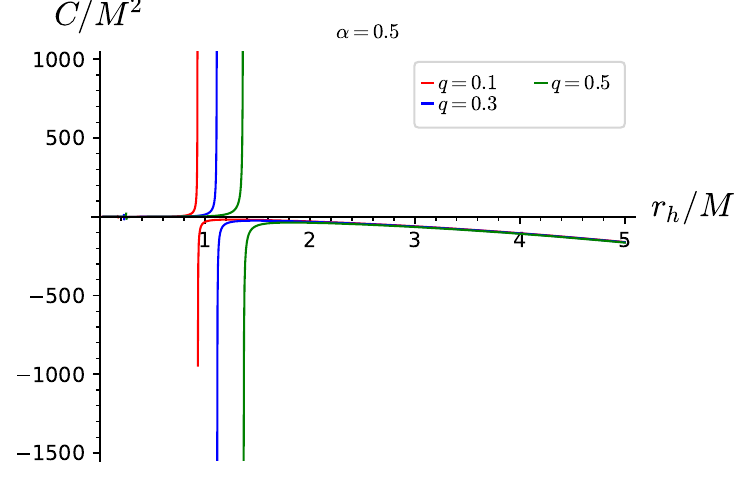} }
    \caption{Variation of Heat capacity with horizon radius for (a) fixed $\alpha_0=0.3$ and (b) $\alpha_0 = 0.5$}
    \label{heat_cap_plots}
    \end{center}
\end{figure}

Heat capacity at a constant pressure is given by,
\begin{equation}\label{eq.29}
    C_p = T\frac{\partial S}{\partial T}.
\end{equation}
\newpage
\noindent Using Eqs. (\ref{eq9}) and (\ref{eq.29}) the heat capacity is calculated as,
\begin{strip}
    \begin{equation}\label{eq14}
    C_p = -\frac{2 \pi  r^2 \left(2 \alpha_0^2 q^2+r^4\right) \left[\alpha_0^2 \left(q^2 \left(1-4 r^2\right)-3 r^4\right)-q^2 r^4+r^6\right]}{\alpha_0^4 \left[q^4 \left(6-8 r^2\right)+6 q^2 r^4\right]+\alpha_0^2 r^4 \left[2 q^4+q^2 \left(7-26 r^2\right)-9 r^4\right]-3 q^2 r^8+r^{10}}.
\end{equation}
\end{strip}
\noindent The graphical representation of the heat capacity is shown in Fig. \ref{heat_cap_plots}. It is well-known fact that the positive heat capacity represents a stable system. However, in case of BH, the negative heat capacity signify the thermal instability with their surroundings. In particular, in the case of regular BH, positive heat capacity is possible in some particular ranges of $r_h$ as indicated in the plot drawn.

The sudden change from positive to negative heat capacity indicates the phase transition from stable to unstable system. It is evident from the plot that after a certain combination of $\alpha_0$ and $q$, Frolov BH becomes thermally unstable.
\subsection{Entropy}
The Bekenstein-Hawking entropy is proportional to the area of event horizon and is given by,
\begin{equation}
    S = \frac{A}{4} = \pi r_+^{2},
\end{equation}
where, $r_+$ is given by Eq. \ref{eq_horizon}. Although not every value of $\alpha_0$ (for a fixed value of $q$) yields a real value of $r_+$, the imaginary part can be ignored as it does not represent the physical horizon. As shown in the graph below, beyond a certain value of $\alpha_0$, $r_+$ becomes complex, leading to an imaginary component in the entropy.
\begin{figure}[H]
     \centering
     {\includegraphics[width=0.45\textwidth]{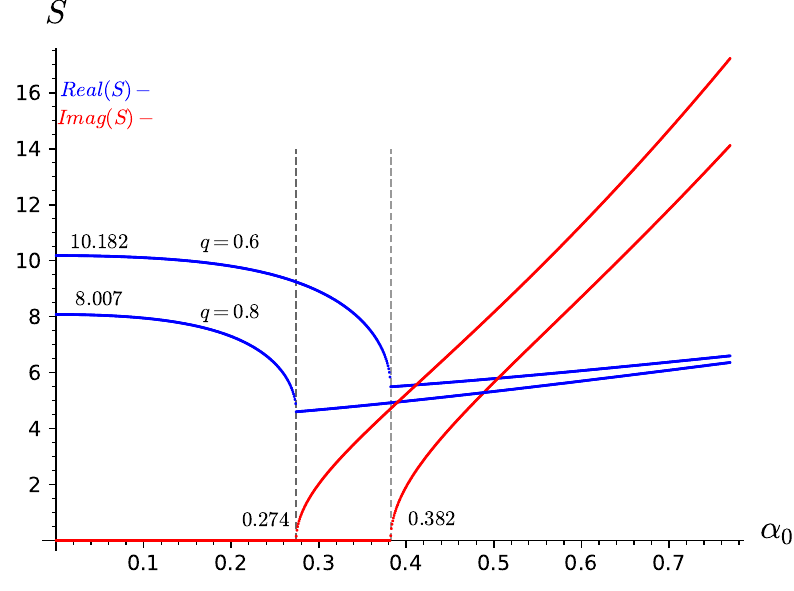} }
     {\includegraphics[width=0.48\textwidth]{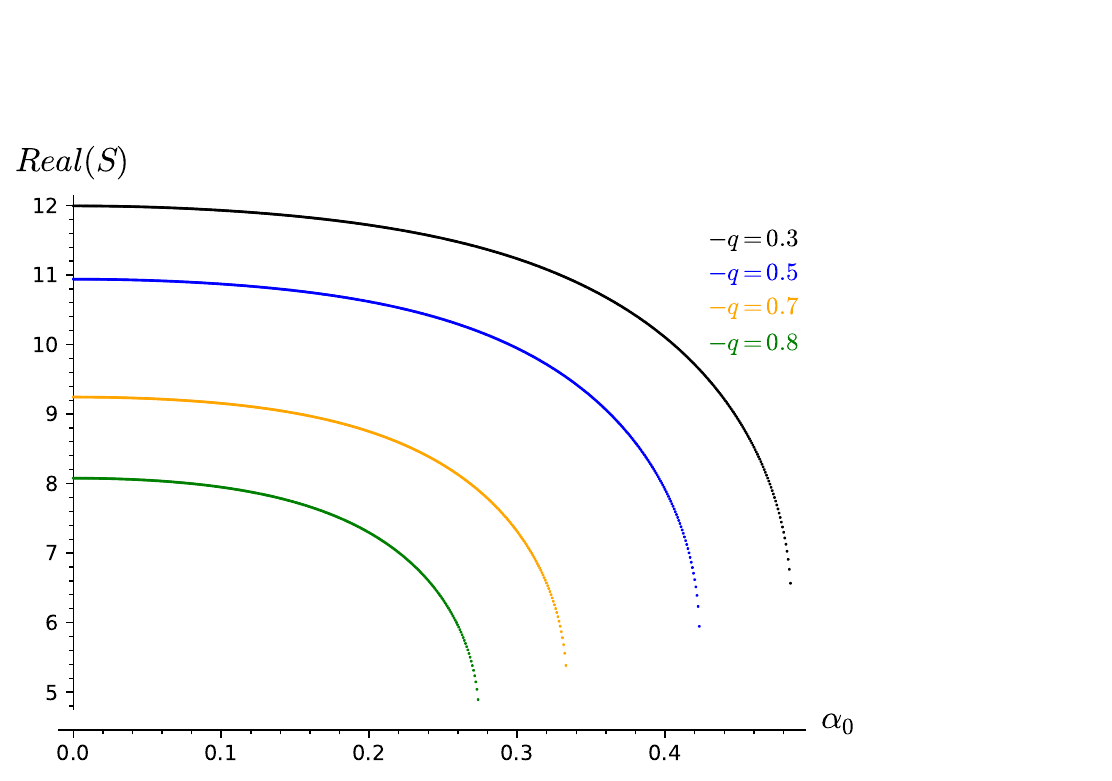} } 
     \caption{Variation of entropy with horizon radius for (a) with imaginary and  real part (b) Considering only real part. Here we have taken $M=1$.} \label{thermal_a_plots}
 \end{figure}
\section{Deflection of Light} \label{section4}
\noindent In the context of GL, a BH can be modeled as a lens positioned at a finite distance between the observer (O) and the source (S), deflecting light rays through its gravitating field which arises due to spacetime curvature. The deflection angle of light in this scenario, can be calculated along the equatorial plane $\theta=\pi/2$ by the formula \cite{Ishihara:2016vdc},
\begin{equation}
   \widehat{\delta}_{angle} = \Psi_{O} - \Psi_{S} + \Phi_{OS}.
\end{equation}
Here, $\Phi_{OS} = \Phi_{O}-\Phi_{S}$, where $\Phi_{O}$ and $\Phi_{S}$ are the angular coordinates of the observer and source respectively. We consider a quadrilateral $^\infty_O \Box ^\infty_S$ of spatial light ray curve from the observer and source, and a circular arc segment $C_{r}$ of coordinate radius $r_{c}(r_{c}\xrightarrow{} \infty)$. One can use the GBT to this quadrilateral and then the deflection angle formula can be reads,  
\begin{equation} \label{daformula}
    \widehat{\delta}_{angle} = - \int \int_{^\infty_O \Box ^\infty_S} \mathcal{K} dS + \int^{O}_{S} k_{g} dl.
\end{equation}
Here, $\mathcal{K}$ and $k_{g}$ denote the Gaussian curvature of the surface of light propagation and the light curve’s geodesic curvature, while $dS$ is the infinitesimal surface area element and dl is the infinitesimal arc line element.
To define the Riemannian manifold for which the light geodesics are interpreted as spatial
curves, we should exploit the null geodesic condition $dS^{2}=0$, which gives,
\begin{equation}
    dt = \pm \sqrt{\gamma_{ij} dx^{i} dx^{j}} + \eta_{\phi} d\phi,
\end{equation}
where $\gamma_{ij}$ is a spatial metric and $\eta_{\phi}$ is the corresponding one-form. In the equatorial plane at constant $t$ of the spacetime metric, one has a 2-dimensional curved space which is represented by,
\begin{equation}
    dl^2 =\gamma_{ij} dx^{i} dx^{j}.
\end{equation} Therefore, the Gaussian curvature corresponding to Riemannian component ($M^{\alpha2}$) can be calculated as,
\begin{equation}
    \int \mathcal{K} \sqrt{\gamma} dr = \int -\frac{\partial}{\partial r} \left( \frac{\sqrt{\gamma}}{\gamma_{rr}} \Gamma^{\phi}_{r\phi}\right) = - \frac{\gamma_{\phi\phi},r}{2\sqrt{\gamma}} \;.
\end{equation} 
As a result, the surface integration of Gaussian curvature given as,
\begin{equation} \label{sintgauc}
    \int \int_{D_{\infty}} \mathcal{K} ds = \int_{\phi_{S}}^{\phi_{R}} \int_{r_{\gamma}}^{\infty} \mathcal{K} \sqrt{\gamma} dr d\phi .
\end{equation}
Here, $r_{\gamma}$ is the radial curvature of geodesics  along the source and observer, $\phi_{S}$ and $\phi_{R}$ are the azimuthal coordinate of source and receiver respectively. $\gamma$ represents the determinant of optical metric. Now, using the formalism, the optical metric corresponding to metric \eqref{eq_metric} is given as,
\begin{equation}
    dl^{2} = \frac{1}{f(r)^{2}} dr^{2} + \frac{r^{2}}{f(r)} d \phi^{2}.
\end{equation}
The Gaussian curvature of the BH is expressed as the form given below, 
\begin{equation} \label{gaucur}
    \begin{aligned}
        \mathcal{K} &= -\frac{2M}{r^{2}}  + \frac{3q^{2}}{r^{4}} + \frac{3M^{2}}{r^{4}} \\
        &\quad - \frac{6Mq^{2}}{r^{5}} -\frac{20q^{2}M^{2}}{r^{6}} \\
        &\quad + \frac{40 M^{2} \alpha_{0}^{2}}{r^{6}} 
        + \mathcal{O} (M^{3}, q^{3}, \alpha_{0}^{4}).
    \end{aligned}
\end{equation}
In this case, the determinant of the optical metric can be expressed as follows,
\begin{equation}
    \gamma = r^{2} \left[ 1 - \left(1-\frac{(2Mr-q^{2})r^{2}}{r^{4}+(2Mr+q^{2})\alpha_{0}^{2}}\right) \right]^{-3}.
\end{equation}
which provides the following needed approximation,
\begin{equation}
    \sqrt{\gamma} \approx r + \mathcal{O} (M^{3}, q^{3}, \alpha_{0}^{4}).
\end{equation}
To keep the order $\mathcal{O} (M^{3}, q^{3}, \ell^{4})$ in deflection angle computations, the computation gives,
\begin{equation}
    \begin{aligned}
        \int_{\phi_{S}}^{\phi_{R}} \int_{r_{0}}^{\infty} \mathcal{K} \sqrt{\alpha_0} \, dr \, d\phi 
        &= -\frac{2M}{r}  + \frac{3q^{2}}{r^{3}} + \frac{3M^{2}}{r^{3}} \\
        &\quad - \frac{6Mq^{2}}{r^{4}} -\frac{20q^{2}M^{2}}{r^{5}} \\
        &\quad + \frac{40 M^{2} \alpha_{0}^{2}}{r^{5}} 
        + \mathcal{O} (M^{3}, q^{3}, \alpha_{0}^{4}) \;.
    \end{aligned}
\end{equation}
where $r_{0}$ is the distance of the closest approach corresponding to the solution of the orbit equation. In order to obtain the deflection angle, the orbit of Eq. \ref{orbiteq} can be expressed as,
\begin{equation}
    \begin{aligned}
        \left( \frac{du}{d\phi}\right)^{2} &= \frac{1}{b^{2}} - u^{2} \bigg( 1 + 2Mu + q^{2}u^{2} \\
        &\quad + 4M\alpha_{0}^{2}u^{4} -\alpha_{0}^{2}q^{4}u^{6} \bigg).
    \end{aligned}
\end{equation}
This equation can be solved by using the perturbative method which provides that,
\begin{equation}
    u(\phi) = \frac{1}{b} \sin{(\phi)} + \mathcal{O} (M).
\end{equation}
In the weak field approximation, the surface integral of Gaussian curvature is given in Eq. \ref{sintgauc} becomes,
\newpage
\begin{strip}
\begin{align}
   - \int_{\phi_{S}}^{\phi_{0}} \int_{r_{0}}^{\infty} \mathcal{K} \sqrt{\alpha_0} \, dr \, d\phi 
   &\approx - \int_{\phi_{S}}^{\phi_{0}} \int_{0}^{u(\phi) = \frac{1}{b} \sin(\phi)} 
   \Bigg[ 2M - 3M^{2}u - 3q^{2}u 
    + 6Mq^{2}u^{3} + 20q^{2}M^{2}u^{4} - 40M^{2}\alpha_{0}^{2}u^{4} \Bigg] du \, d\phi \;, \notag\\
   &\approx  \frac{2M}{b} \left(\sqrt{1-b^{2}u_{0}^{2}} + \sqrt{1-b^{2}u_{s}^{2}}\right) \notag\\
   &\quad -\frac{Mq^{2}}{3b^{3}} \Bigg[ (16+b^{2}u_{0}^{2})\sqrt{1-b^{2}u_{0}^{2}} 
   + (16+b^{2}u_{s}^{2})\sqrt{1-b^{2}u_{s}^{2}} \Bigg] \notag \\
   &\quad -\left( \frac{3q^{2}}{4b} + \frac{M^{2}}{4b} \right) 
   \left( u_{0}\sqrt{1-b^{2}u_{0}^{2}} + u_{s}\sqrt{1-b^{2}u_{s}^{2}} \right) \notag \\
   &\quad -\left( \frac{3q^{2}}{4b^{2}} - \frac{15M^{2}}{4b^{2}} 
   + \frac{15M^{2}\alpha_{0}^{2}}{4b^{4}} - \frac{27M^{2}q^{2}}{64b^{4}} \right) 
   (\arccos{bu_{0}} + \arccos{bu_{s}}) \notag \\
   &\quad -\frac{5M^{2}\alpha_{0}^{2}}{4b^{3}} 
   \Bigg[ u_{0}(3+2b^{2}u_{0}^{2}) \sqrt{1-b^{2}u_{0}^{2}} 
   + u_{s} (3+2b^{2}u_{s}^{2}) \sqrt{1-b^{2}u_{s}^{2}} \Bigg]. 
\end{align}
\end{strip}
\begin{figure}[H] 
	\begin{center}
        {\includegraphics[height=5cm, width=0.4\textwidth]{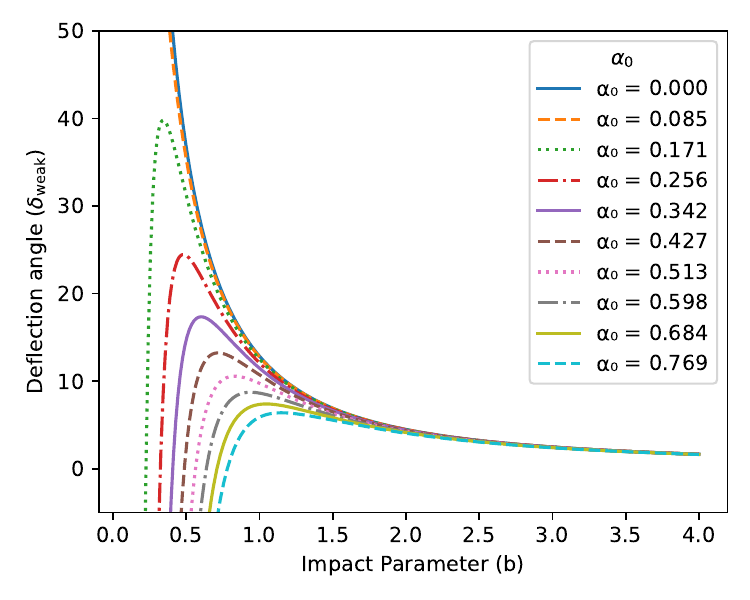}} 
        {\includegraphics[height=5cm, width=0.4\textwidth]{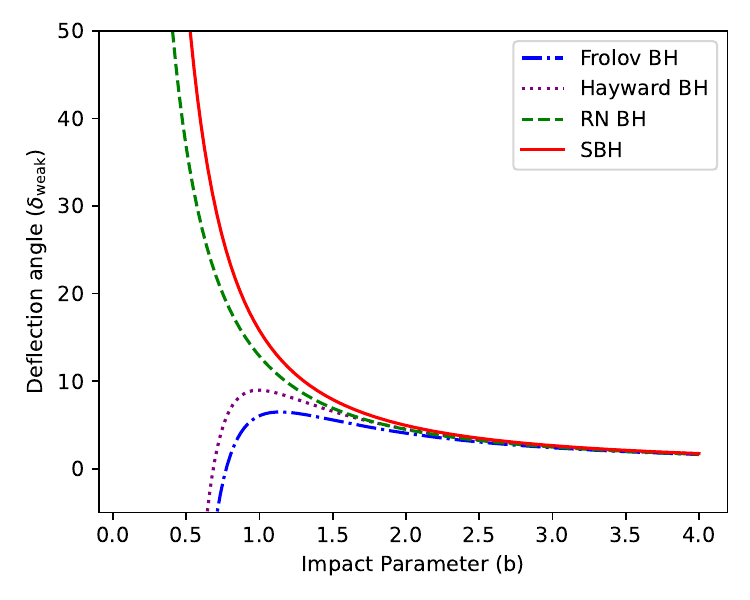}}
	\end{center}
	\caption{The variation of deflection angle with impact parameter for different values of $\alpha_0$ parameter (Here we consider q=0.5). For comparison, deflection angle of well-known BHs to which the Frolov BH reduces under specific limiting cases are also illustrated. If the values of $q$ and $\alpha_0$ are not explicitly mentioned, both parameters are assumed to be 0.5 for consistency.} \label{def}
\end{figure}
Here, $u_{O}$ and $u_{S}$ are the reciprocal of the observer and source distance from the BH respectively and we use $\cos\phi_{O}=-\sqrt{1-b^{2}u_{O}^{2}}$ and $\cos\phi_{S}=-\sqrt{1-b^{2}u_{S}^{2}}$. The geodesic curvature vanishes for a non-rotating BH and, therefore, makes no contribution in the deflection angle. The above integral can apparently diverge in the limit $\ bu_{O} \xrightarrow{} 0$ and $\ bu_{S} \xrightarrow{} 0$. This divergence could be linked to the Hubble length which is related to an effective cosmological constant. In the weak field approximation, the deflection angle can be calculated using Eq. \ref{daformula} and expressed as,
\begin{equation}
    \begin{aligned}
        \delta_{weak} &= \frac{4M}{b} + \frac{15\pi M^{2}}{4b^{2}} -\frac{3 \pi q^{2}}{4 b^{2}} -\frac{32 M q^{2}}{3b^{3}} \\
        &\quad + \frac{27 \pi M^{2} q^{2}}{64b^{4}} -\frac{15 \pi M^{2} \alpha_{0}^{2}}{4 b^{4}}.
    \end{aligned}
\end{equation}
The expression obtained for the deflection angle reduces to the well-known Schwarzschild, Hayward, and RN BHs in the prescribed limits. The last term in the expression, which involves the Hubble length and a negative sign, indicates a decrease in the deflection angle. The graphical representation of the deflection angle is shown in Fig. \ref{def}. It is clearly observed that the deflection angle decreases with an increase in the Hubble length parameter. Physically, this suggests that as the Hubble length increases, the effect of cosmic expansion becomes more prominent, leading to a weaker gravitational lensing effect around the BH. This indicates that the presence of the Hubble length parameter reduces the bending of light near the BH, reflecting the influence of the expanding universe on gravitational phenomena.
\section{Black Hole Shadows} \label{section5}
\subsection{BH Shadow Radius and Parameter Constraints with Sgr A*}
\begin{figure}[H] 
	\begin{center}
        {\includegraphics[height=5cm, width=0.35\textwidth]{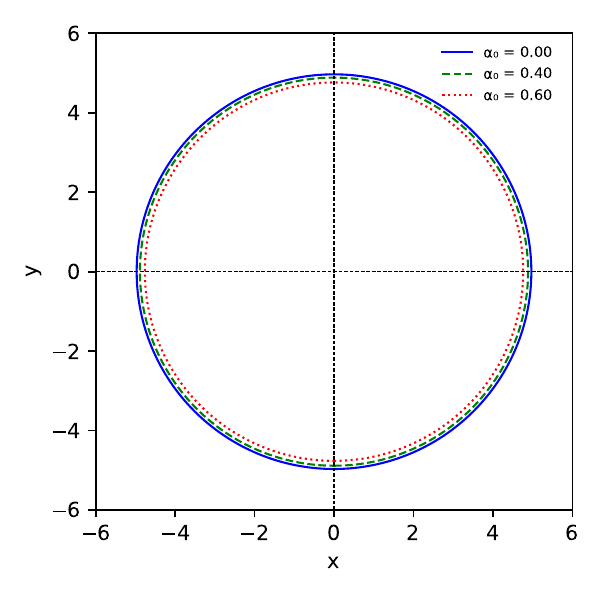}} 
        {\includegraphics[height=5cm, width=0.35\textwidth]{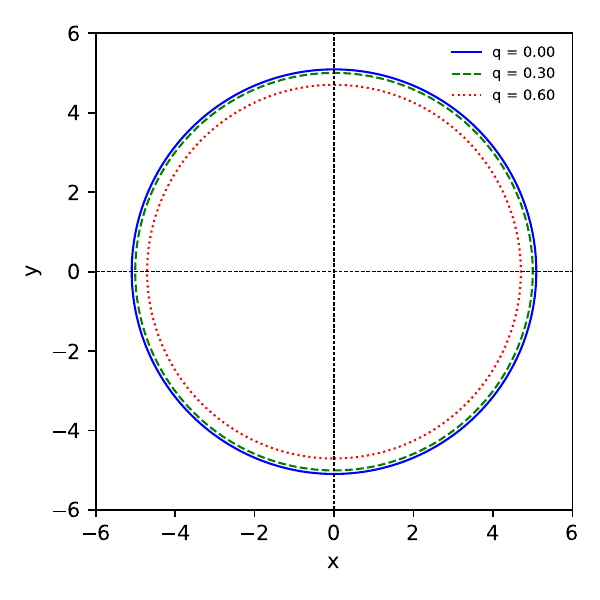}}
        {\includegraphics[height=5cm, width=0.35\textwidth]{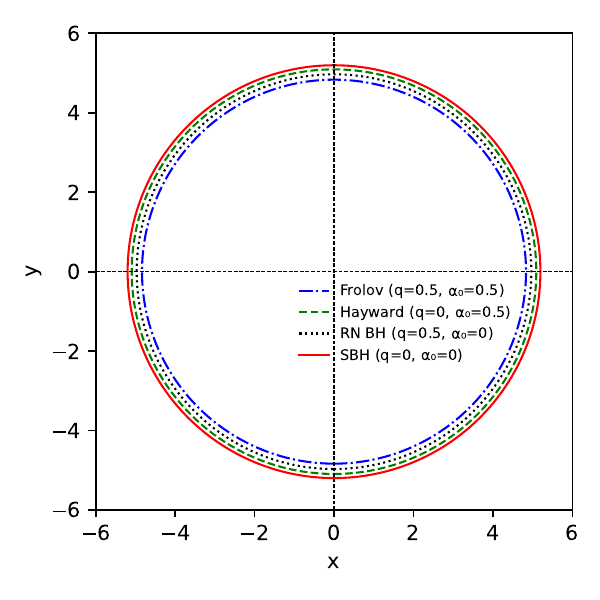}}
        {\includegraphics[height=5cm, width=0.35\textwidth]{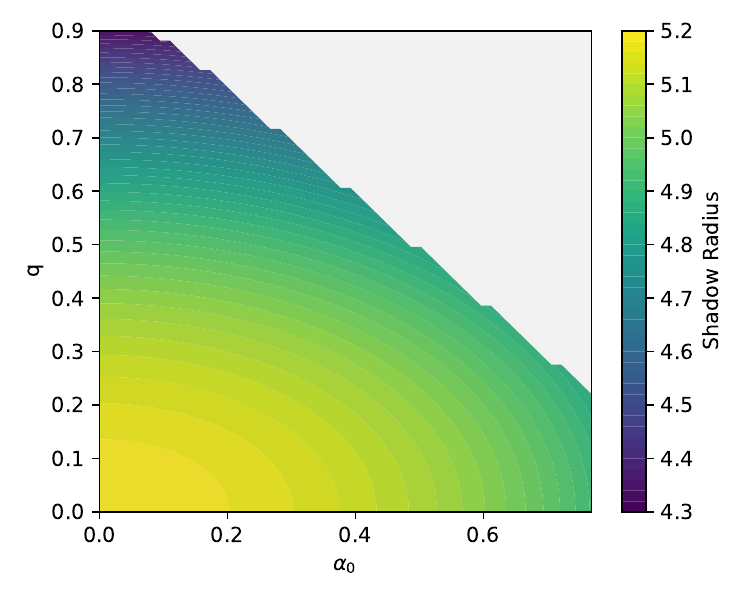}}
	\end{center}
	\caption{The variation of BH shadow of Frolov BH and its dependency on the parameters $q$ and $\alpha_{0}$. The comparison of shadow of Frolov BH with other well-known BH solution in GR. The two-dimensional contour plot illustrates the variation of the BH shadow radius as a function of both parameters.} \label{radiusSH}
\end{figure}

In this section we investigate the radius of BH shadow following the procedure of Perlick and Tsupko \cite{Perlick:2021aok}. In the class of spherically symmetric and static metrics, those possessing a photon sphere are precisely the ones where an observer can see an infinite number of images of a light source. Therefore, we can calculate the photon orbit radius through the following relation:
\begin{equation}
    f'(r_{p})r_{p}-2 f(r_{p}) = 0,
\end{equation}
which gives,
\begin{align}
    &\frac{2 r_p^2 (2M r_p - q^2) (r_p^4 - (2M r_p + q^2) \alpha_0^2)}
    {(r_p^4 + (2M r_p + q^2) \alpha_0^2)^2} \notag \\
    &\quad + \frac{2 (2M r_p - q^2) r_p^2}{r_p^4 + (2M r_p + q^2) \alpha_0^2} = 0.
\end{align}
 This equation is complex and does not admit an analytical solution. To solve this equation, we employ a numerical approach to determine the radii of the photon orbit $r_{p}$. Once the photon orbit radii are obtained, the shadow radii can be computed using the following relation:
\begin{equation} \label{shadpwradiuseq}
    R_{sh}= \frac{r_{p}}{\sqrt{f(r)}|_{r=r_{p}}}.
\end{equation}
The variation of the shadow radius with $\alpha_{0}$ and $q$ is illustrated in Fig. \ref{radiusSH}, along with a comparison of the Frolov BH with other well-known BH solutions in GR. Our results indicate that as both $\alpha_{0}$ and $q$ increase, the shadow radius decreases. Furthermore, among the BHs considered, the Schwarzschild BH exhibits the largest shadow radius, followed by the Hayward BH, then the RN BH, with the FroloV BH having the smallest shadow radius. The reduction in the shadow radius due to the parameters $\alpha_{0}$ and $q$ suggests that both Hubble length and charge parameter, which is associated with modifications to the spacetime structure, effectively weaken the GL effect. This implies that the presence of these parameters reduces the effective gravitational influence of the BH, leading to a smaller apparent shadow size.
The EHT collaboration has analyzed uncertainties in Sgr A*’s shadow size using mass-to-distance measurements from Keck and VLTI \cite{Do:2019txf,GRAVITY:2020gka}. They introduced a fractional deviation parameter, $\delta$, which quantifies how much the observed shadow radius deviates from the Schwarzschild BH prediction, $r_{\text{sh, Schwarzschild}} = 3\sqrt{3} M$. The Keck and VLTI estimates give the following \cite{akiyama2022L17}:
\begin{itemize}
    \item Keck: $\delta = -0.04^{+0.09}_{-0.10}$,
    \item VLTI: $\delta = -0.08^{+0.09}_{-0.09}$.
\end{itemize}
\begin{figure}[H] 
	\begin{center}
        {\includegraphics[width=0.48\textwidth]{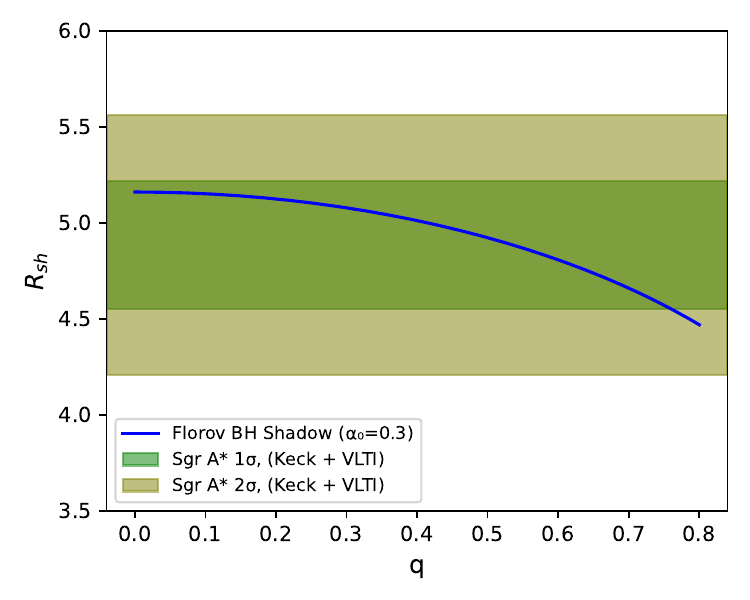}} 
        {\includegraphics[width=0.48\textwidth]{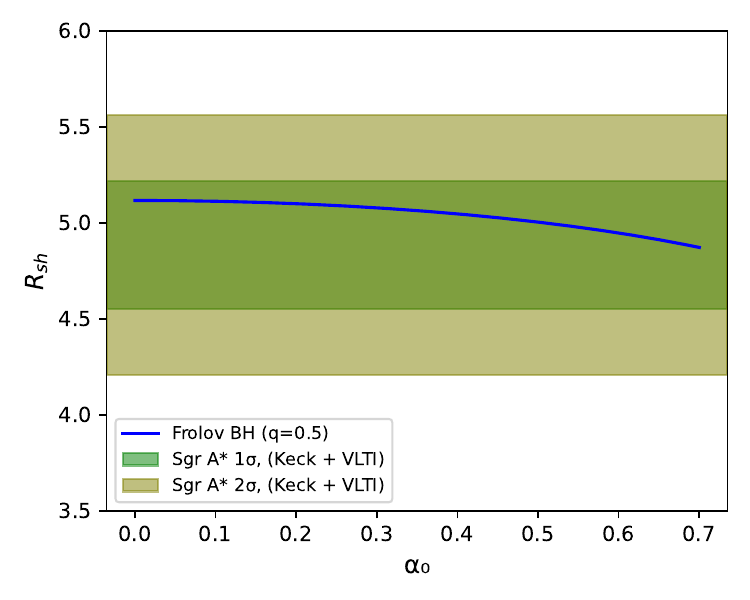}}
	\end{center}
	\caption{Shadow radius of Frolov BH (solid blue line) as a function of $q$ (upper panel) and as a function of $\alpha_{0}$ (lower panel). The dark green and olive green regions are consistent with the EHT horizon-scale image of Sgr A* at $1\sigma$ and $2\sigma$ respectively, after averaging the Keck and VLTI mass-to-distance ratio priors for Sgr A*. } \label{SHEHT12}
\end{figure}
Averaging these values results in $\delta \simeq -0.060 \pm 0.065$, which translates into constraints on the shadow radius given follows:
\begin{equation}
4.55 \lesssim \frac{r_{\text{sh}}}{M} \lesssim 5.22 \quad (1\sigma),
\end{equation}
\begin{equation}
4.21 \lesssim \frac{r_{\text{sh}}}{M} \lesssim 5.56 \quad (2\sigma).
\end{equation}
These constraints, slightly more precise than previous estimates, suggest a marginally smaller shadow than the Schwarzschild prediction. We will use these bounds to test our BH shadow model.
\begin{figure}[H] 
	\begin{center}
        {\includegraphics[width=0.48\textwidth]{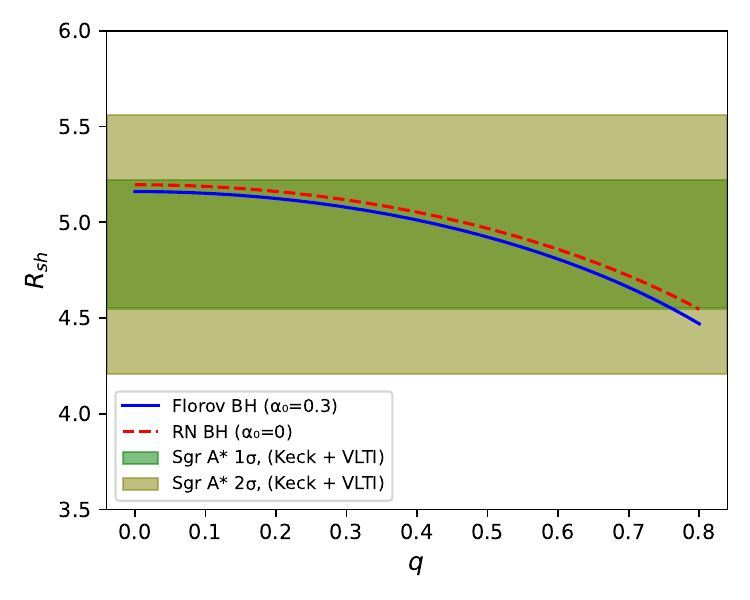}} 
        {\includegraphics[width=0.48\textwidth]{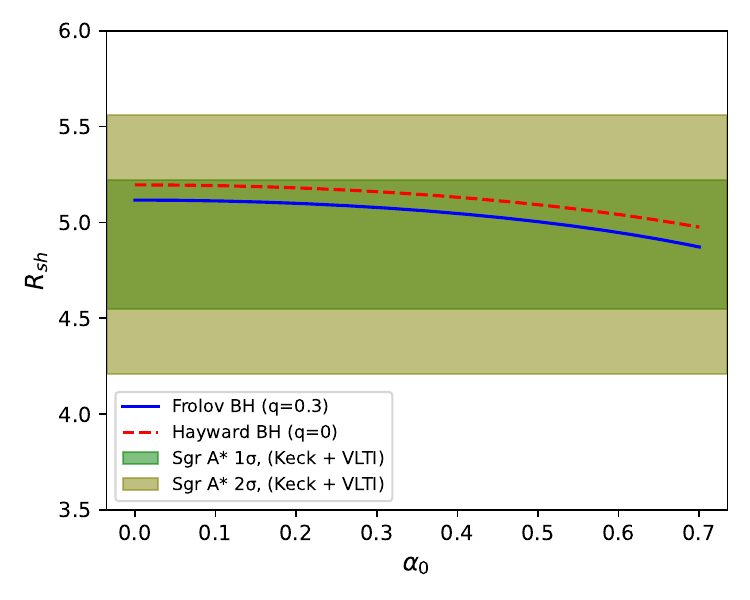}}
	\end{center} 
	\caption{Shadow radius of Frolov BH (solid blue line) and RN BH (red dashed line) as a function of $q$ (upper panel). Shadow radius of Frolov BH (solid blue line) and Hayward BH (red dashed line) as a function of $\alpha_{0}$ (lower panel).} \label{SHEHT34}
\end{figure}
\noindent The study examines the constraints on the charge while fixing $\alpha_{0}$ or vice-versa. It is observed from Fig. \ref{SHEHT12}, the shadow radius varies slowly with $\ell$, compared to $q$. Numerical calculations show that for $\alpha_{0} = 0.3$, the EHT observations set upper bounds of $q \lesssim 0.8M$ (1$\sigma$) and $q \lesssim 0.9M$ (2$\sigma$). However, for $q=0.3$, the numerically calculated shadow radius remains consistent with the EHT observations, indicating that Sgr A* can be interpreted as a Frolov BH for any allowed value of the Hubble length parameter. These constraints are comparable to those for the Hayward and RN BH metric as result is depicted in Fig. \ref{SHEHT34}, and are consistent with the study presented by Vagnozzi et al. \cite{Vagnozzi:2022moj}. Furthermore, increasing $\alpha_{0}$ slightly decreases the size of the shadow, suggesting a negative correlation between $\alpha_{0}$ and $q$.
\subsection{BH Shadow Images with Spherical Accretion Flow}
Here, we consider a simple model of an optically thin spherical accretion flow around a regular Frolov BH. The accreting matter undergoes free fall from infinity toward the BH. This infalling model is considered more realistic than the static accretion model, as most accreting matter in the universe is inherently dynamic \cite{Falcke:1999pj}. Here, we refer to some selected studies conducted in this area \cite{Moffat:2019uxp,Zeng:2021dlj,Guerrero:2021ues,Shaikh:2021cvl,He:2021htq,Zare:2024dtf}. For emission mechanisms, certain assumptions shall be made for the calculation of the intensity from the radiating accretion flow.
The observed specific intensity $I_{\nu 0}$ at the observed photon frequency $\nu_\text{obs}$ at the point $(X,Y)$ of the observer's image (usually measured in $\text{erg} \text{s}^{-1} \text{cm}^{-2} \text{str}^{-1} \text{Hz}^{-1}$) is given by \cite{jaroszynski1997optics},
\begin{eqnarray} \label{intensityintegraleq}
    I_{obs}(\nu_{obs},X,Y) = \int_{\gamma}g^3 j(\nu_{e})dl_\text{prop},  
\end{eqnarray}
where $g = \nu_{obs}/\nu_{e}$ is the redshift factor, $\nu_{e}$ is the photon frequency as measured in the rest-frame of the emitter, $j(\nu_{e})$ is the emissivity per unit volume in the rest-frame of the emitter, and $dl_\text{prop} = k_{\rho}u^{\rho}_{e} d\lambda$ is the infinitesimal proper length as measured in the rest-frame of the emitter. The redshift function $g$ can be calculated by the relation \cite{Bambi:2013nla},
\begin{equation}
    g = \frac{k_{\rho}u^{\rho}_{\text{obs}}}{k_{\sigma}u^{\sigma}_{e}}\;.
\end{equation}
Here $k^{\mu}$ is the 4-velocity of the photons, $u^{\rho}_{e}$ 4-velocity of the accreting gas emitting the radiation and  $u^{\mu}_{\text{obs}}$ is 4-velocity of the observer with $\lambda$ being the affine parameter along the photon path $\gamma$. The integral presented in Eq. \ref{intensityintegraleq} should be evaluated along the photon path $\gamma$, corresponding to null geodesics. To generalize the formalism for both cases, we will restrict our analysis to the equatorial plane and assume that the gas is undergoing radial free fall, characterized by a four-velocity with components defined as follows:
\begin{eqnarray}
u^t_{e} & = & \frac{1}{g_{tt}(r)}, \nonumber \\
u^r_{e} & = & -\sqrt{\frac{1-g_{tt}(r)}{g_{tt}(r)g_{rr}(r)}}, \nonumber \\
u^{\theta}_{e} & = & 0, \nonumber \\
u^{\phi}_{e} & = & 0, 
\end{eqnarray}
where the metric components $g_\textit{tt}$, $g_\textit{rr}$ and $g_{\phi\phi}$ are corresponding to the line element given in metric. 
In previous section, the components of four-velocity were already calculated. In order to ease further calculations, we obtain an equation between the radial and time component of the four-velocity as follows,
\begin{figure}[H] 
	\begin{center}
       \includegraphics[width=0.5\textwidth, height=0.6\textheight]{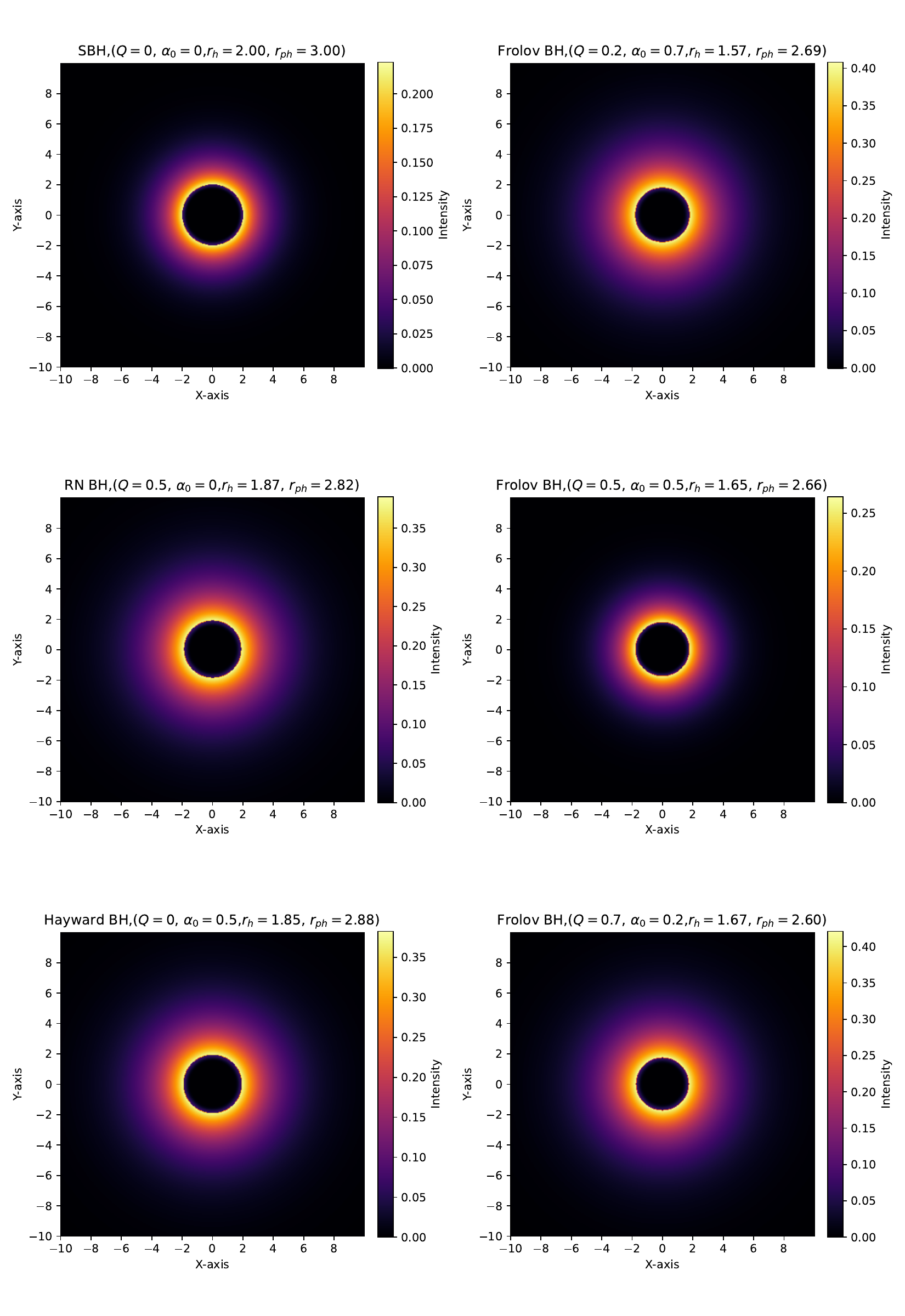}
	\end{center}
	\caption{BH shadow with infalling spherical accretion. The left panel illustrates the shadow images of several well-known BHs, while the right panel shows the shadow images of Frolov BH for different values of $q$ and $\alpha_{0}$.} \label{SHAcreetionFlow}
\end{figure}
\begin{equation}
    \frac{k_r}{k_t} = \pm \sqrt{g_{rr}\bigg(\frac{1}{g_{tt}}-\frac{b^2}{g_{\phi\phi}}\bigg)},
\end{equation}
where the sign $ +(-) $ corresponds to photon approaching (going
away) from the massive object.
The redshift function $g$ is then given by,
\begin{eqnarray}
   g =\frac{1}{\frac{1}{g_{tt}} \pm \frac{k_r}{k_t}\sqrt{\bigg(\frac{1-g_{tt}}{g_{tt}g_{rr}}\bigg)}}.
\end{eqnarray}
The specific emissivity for a simplified model, assuming monochromatic emission at the emitter's rest-frame frequency $\nu_{\star}$, and a radial dependence that decreases with distance as $1/r^2$, can be expressed as \cite{Gralla:2019xty},
\begin{equation}
    j(\nu_{e}) \propto \frac{\delta_D(\nu_{e}-\nu_{\star})}{r^2},
\end{equation}
where $\delta_D$ is the Dirac delta function. The proper length can be written as,
\begin{equation}
    dl_{\text{prop}} = k_{\rho}u^{\rho}_{e}d\lambda = -\frac{k_t}{g|k^r|}dr\;.
\end{equation}
By integrating the intensity over all observed frequencies, we obtain the expression for the observed photon intensity as follows,
\begin{equation}
    I_{obs}(X,Y) \propto -\int_{\gamma} \frac{g^3 k_t}{r^2k^r}dr.
\end{equation}
We are now well-positioned to proceed with the computation of the corresponding images. This will allow us to visualize the BH images and analyze its key features in detail. We closely follow the numerical technique presented in \cite{Saurabh:2020zqg} and modified for the use of Frolov BH. In the Fig. \ref{SHAcreetionFlow}, we present the optical appearance of the specific intensity of the Frolov BH surrounded by a thin spherical accretion disk. For comparison, we also include the intensity profiles of other well-known BH solutions to highlight the differences. Our results show that the intensity increases with an increasing Hubble length parameter. Furthermore, when the charge parameter is coupled with the Hubble parameter, as in the case of the Frolov BH, the intensity reaches its maximum for the minimum charge value. These results indicate that BHs with different $\alpha_{0}$ and $q$ values could have distinct observational signatures, which might be relevant for identifying exotic BH solutions in astrophysical data.
\section{Energy Emission Rate} \label{section6}
In this section, we study the energy emission rate (EER) of a Frolov regular BH. The expression of EER reads as \cite{Wei:2013kza},
\begin{equation}
	\frac{d^{2} Z(\omega)}{d\omega dt} = \frac{2 \pi^{2} \sigma_{lim}}{\exp(\frac{\omega}{T_{H}})-1} \omega^{3},
\end{equation}
where the parameters $Z(\omega)$, $\omega$ and $T_{H}$ represent the energy, frequency and Hawking temperature respectively corresponding to the BH. The expression of limiting constant value $\sigma_{lim}$ for this particular regular BH can be expressed as \cite{Decanini:2011xw},
\begin{equation}
	\sigma_{lim} = \pi 	R_{sh}^{2}\;.
\end{equation}
\begin{figure}[H] 
	\begin{center}
        {\includegraphics[width=0.35\textwidth]{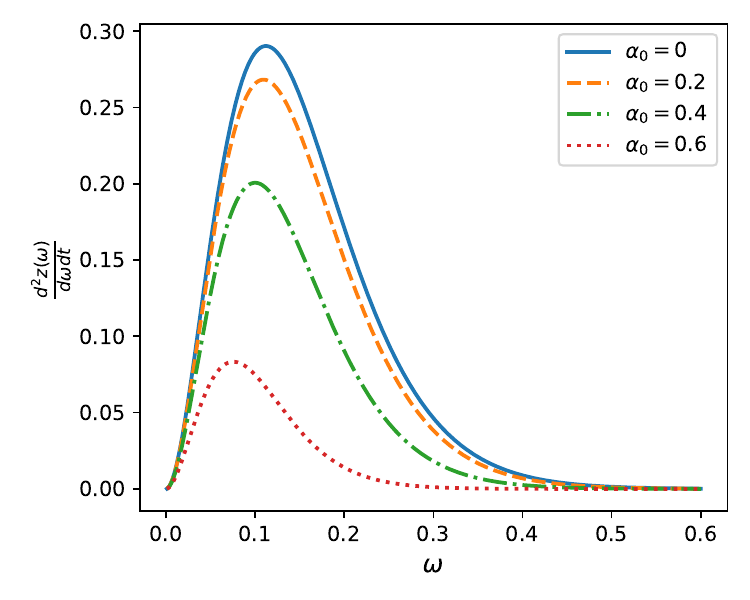}} 
        {\includegraphics[width=0.35\textwidth]{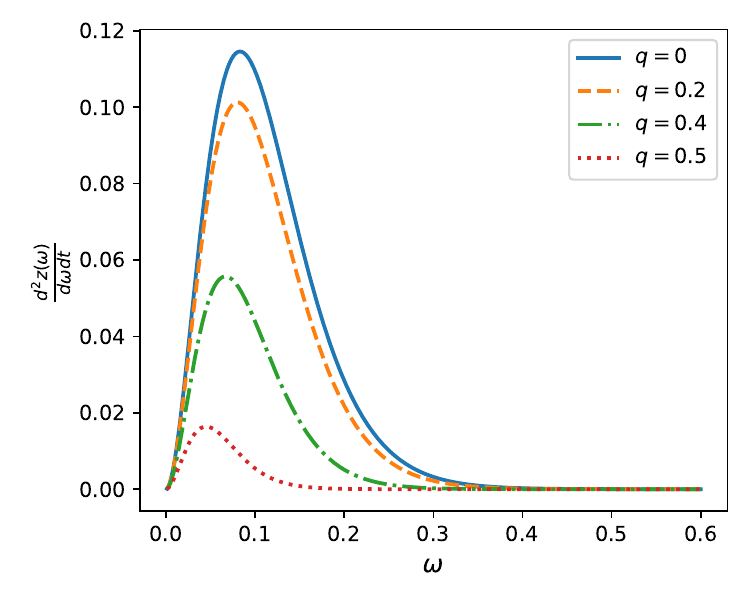}}
        {\includegraphics[width=0.35\textwidth]{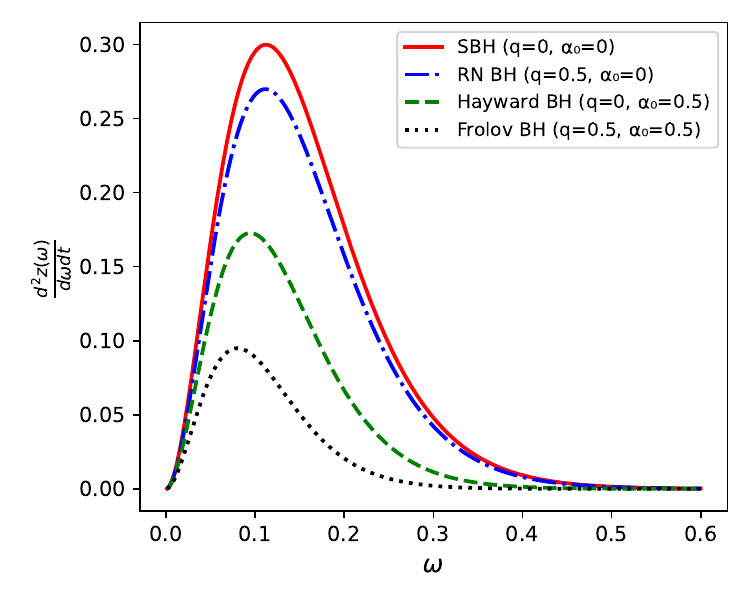}}
	\end{center}
	\caption{The variation of EMR of Frolov BH and its dependency on the parameters $q$ and $\alpha_{0}$. The comparison of shadow of Frolov BH with other well-known BH solution in GR.} \label{EERfig}
\end{figure}
\noindent Here, $R_{sh}$ denotes the shadow radius of the BH, as previously obtained using the expression provided in Eq. \ref{shadpwradiuseq}. Therefore, the expression of EER for this BH spacetime becomes,
\begin{equation}
	\frac{d^{2} Z(\omega)}{d\omega dt} = \frac{2 \pi^{3} 	R_{sh}^{2}} {\exp(\frac{\omega}{T_{H}})-1} \omega^{3}.
\end{equation} 
The variation of EERs as a function of frequency is depicted in Fig. \ref{EERfig}. The EERs of Frolov BH increase as the charge $\alpha_{0}$ and the Hubble length parameter $\alpha_{0}$ decrease. The maximum EER occurs at $\alpha_{0}$ and $q=0$, corresponding to a Schwarzschild BH. The presence of charge and $\alpha_{0}$ suppresses the EER, suggesting that these parameters introduce factors that dilute the energy available for emission. Specifically, $\alpha_{0}$, linked to cosmological scales, and charge $q$ probably influence the interaction of BH with its environment, reducing the emission of energy.

\section{Conclusion} \label{section7}
In this paper, we studied the properties of null geodesics, thermodynamics, gravitational lensing, and BH shadows in the vicinity of a static regular Frolov BH. We analyzed light bending in the weak field regime, constrained BH shadow using EHT collaboration data, and examined shadow images with spherically symmetric accretion flow along with the energy emission rate. Our results reveal distinct differences between the Frolov BH and well-known solutions like Schwarzschild, RN, and Hayward BHs. The main findings can be summarized as follows:
\begin{itemize}[label=$\bullet$]
    \item The Frolov BH can possess up to two horizons, depending on the appropriate choice of parameters. The allowed parameter space $(q, \alpha_0)$ for the existence of BH horizons becomes increasingly compact as the charge $q$ increases.
    \item The effective potential exhibits minima, indicating the possibility of stable circular orbits, while its peaks correspond to the radii of unstable photon spheres. Ray tracing of null geodesics clearly illustrates the trajectories of massless particles around the Frolov BH. The study of the stability of null geodesics suggests that the Frolov BH exhibits the most unstable photon orbits, while the Schwarzschild BH has the least. This indicates that the charge $(q)$ and the Hubble length parameter $(\alpha_{0})$ enhance the susceptibility of photon orbits to perturbations, increasing their instability.
    \item The Hawking temperature reaches its maximum at a low horizon radius, peaking at a specific value before gradually decreasing and eventually stabilizing. Additionally, an increase in the charge parameter leads to a decrease in the Hawking temperature. We also observe that the Frolov BH exhibits positive heat capacity within a specific range of the horizon radius, followed by a sudden transition to negative heat capacity, indicating a phase transition from a stable to an unstable state.
    \item The deflection angle decreases with an increase in the Hubble length parameter. The nature of Schwar-zschild and RN BHs is distinctly different from that of Hayward and Frolov BHs, suggesting that the negative deflection observed in Hayward and Frolov BHs arises solely due to the presence of the Hubble length parameter. Physically, this implies that the influence of cosmic expansion, encoded in the Hubble length parameter, alters the spacetime curvature, leading to a reduction of the GL effect compared to traditional BH solutions.
    \item The shadow radius decreases with increasing $\alpha_0$ and $q$, with the Schwarzschild BH exhibiting the largest shadow and the Frolov BH the smallest. This indicates that the Hubble length and charge parameters, by modifying the spacetime geometry, reduce the effective deflection of light rays, leading to a smaller apparent shadow size. The shadow radius is more sensitive to $q$ than $\alpha_{0}$, with EHT data constraining $q$ but allowing Sgr A* to be interpreted as a Frolov BH for any permissible value of $\alpha_{0}$.
    \item The specific intensity of the Frolov BH increases with $\alpha_{0}$ and peaks at minimal $q$ values. Comparing with other BHs reveals distinct observational signatures, offering insights into identifying exotic astrophysical BHs surrounded by the spherical accreting flow.
    \item The EER of the Frolov BH increases as the charge $q$ and Hubble length parameter $\alpha_{0}$ decrease, reaching a maximum at $q = 0$ and $\alpha_{0} = 0$, corresponding to the Schwarzschild BH. The presence of $q$ and $\alpha_{0}$ suppresses EER, probably due to their influence on the interaction of BH with its environment, reducing energy emission.
\end{itemize}
The future extension of this study will investigate the influence of homogeneous and non-homogeneous plasma profiles on radiation propagation and gravitational lensing, which is essential for accurately modeling astrophysical observations of compact objects.
\newpage
\section{Acknowledgments}
The authors SK and HN acknowledge the use of facilities at ICARD, Gurukula Kangri (Deemed to be University), Haridwar, India. The author, SK, sincerely acknowledges IMSc for providing exceptional research facilities and a conducive environment that facilitated his work as an Institute Postdoctoral Fellow. One of the authors, H.N., would also like to thank IUCAA, Pune, for the support under its associateship program where a part of this work was done. The author HN also acknowledges the financial support provided by the Science and Engineering Research Board (SERB), New Delhi, through grant number CRG/2023/008980.
\bibliographystyle{spphys}
\bibliography{frolov}
\end{document}